\documentclass[twoside,american]{IEEEtran}
\usepackage[T1]{fontenc}
\usepackage[latin9]{inputenc}
\usepackage{color}
\usepackage{array}
\usepackage{float}
\usepackage{bm}
\usepackage{multirow}
\usepackage{amsmath}
\usepackage{amsthm}
\usepackage{amssymb}
\usepackage{graphicx}

\makeatletter

\newcommand{\noun}[1]{\textsc{#1}}
\providecommand{\tabularnewline}{\\}
\floatstyle{ruled}
\newfloat{algorithm}{tbp}{loa}
\providecommand{\algorithmname}{Algorithm}
\floatname{algorithm}{\protect\algorithmname}

\theoremstyle{plain}
\newtheorem{thm}{\protect\theoremname}
\theoremstyle{remark}
\newtheorem{rem}[thm]{\protect\remarkname}
\theoremstyle{plain}
\newtheorem{lem}[thm]{\protect\lemmaname}

\usepackage{graphicx, amsmath}
\usepackage{subfigure}
\usepackage{cite}
\usepackage{algorithmic}
\newcommand{\re}{\text{r}}

\newcommand{\algnew}{3}

\renewcommand{\fnum@figure}{Fig.~\thefigure}
\allowdisplaybreaks

\let\rem\@undefined
\theoremstyle{remark}
\newtheorem{rem}{\protect\remarkname}

\let\lem\@undefined
\theoremstyle{lemma}
\newtheorem{lem}{\protect\lemmaname}

\let\prop\@undefined
\theoremstyle{proposition}
\newtheorem{prop}{\protect\propositionname}

\let\cla\@undefined
\theoremstyle{claim}
\newtheorem{cla}{\protect\claimname}

\makeatother

\usepackage{babel}
\providecommand{\lemmaname}{Lemma}
\providecommand{\remarkname}{Remark}
\providecommand{\theoremname}{Theorem}

\begin{document}

\title{Energy Efficiency Maximization for C-RANs: Discrete Monotonic Optimization,
Penalty, and $\ell_{0}$-Approximation Methods}

\author{Kien-Giang Nguyen,~\IEEEmembership{Student Member, IEEE}, Quang-Doanh
Vu,~\IEEEmembership{Member, IEEE} Markku Juntti,~\IEEEmembership{Senior Member, IEEE},
and Le-Nam Tran,~\IEEEmembership{Senior Member, IEEE}\thanks{This work has been financially supported  by Academy of Finland under
the projects \textquotedblleft Wireless Connectivity for Internet
of Everything\textendash Energy Efficient Transceiver and System Design
(WiConIE)\textquotedblright{} (grant 297803), \textquotedblleft Flexible
Uplink-Downlink Resource Management for Energy and Spectral Efficiency
Enhancing in Future Wireless Networks (FURMESFuN)\textquotedblright{}
(grant 31089), and ``6Genesis Flagship'' (grant 318927). This paper
was presented in part at IEEE International Conference on Acoustics,
Speech and Signal Processing (ICASSP 2017), Alberta, Canada, April
15-20, 2018.}\thanks{K.-G. Nguyen, Q.-D. Vu, and M. Juntti are with the Centre for Wireless
Communications, University of Oulu, P.O.Box 4500, FI-90014, Oulu,
Finland. Email: \{giang.nguyen, doanh.vu, markku.juntti\}@oulu.fi.}\thanks{L.-N. Tran is with School of Electrical and Electronic Engineering,
University College Dublin, Ireland. Email: nam.tran@ucd.ie).}}
\maketitle
\begin{abstract}
We study downlink of multiantenna cloud radio access networks (C-RANs)
with finite-capacity fronthaul links. The aim is to propose joint
designs of beamforming and remote radio head (RRH)-user association,
subject to constraints on users' quality-of-service, limited capacity
of fronthaul links and transmit power, to maximize the system energy
efficiency. To cope with the limited-capacity fronthaul we consider
the problem of RRH-user association to select a subset of users that
can be served by each RRH. Moreover, different to the conventional
power consumption models, we take into account the dependence of baseband
signal processing power on the data rate, as well as the dynamics
of the efficiency of power amplifiers. The considered problem leads
to a mixed binary integer program (MBIP) which is difficult to solve.
Our first contribution is to derive a globally optimal solution for
the considered problem by customizing a discrete branch-reduce-and-bound
(DBRB) approach. Since the global optimization method requires a high
computational effort, we further propose two suboptimal solutions
able to achieve the near optimal performance but with much reduced
complexity. To this end, we transform the design problem into continuous
(but inherently nonconvex) programs by two approaches: penalty and
$\ell_{0}$-approximation methods. These resulting continuous nonconvex
problems are then solved by the successive convex approximation framework.
Numerical results are provided to evaluate the effectiveness of the
proposed approaches.
\end{abstract}


\section{Introduction}

\emph{Coordinated multipoint joint transmission} (CoMP-JT) \cite{marsch2011coordinated}
has been proposed in the current LTE standards to deal with the inter-cell
interference, which is one of the key factors limiting the capacity
of modern wireless communications systems. The central ideal of CoMP-JT
is to allow for joint processing of the data symbols by multiple transmitters,
thereby exploiting the cooperative gains efficiently. Thus, CoMP-JT
is expected to improve the system performance significantly, especially
for the cell-edge users. However, CoMP-JT requires a low-latency and
high-capacity backhaul network, and a strict synchronization mechanism
among transmitters \cite{3gppCoMP}. These requirements are hard to
implement in practice.

\emph{Cloud radio access networks} (C-RANs) are emerging as a revolutionary
solution that can deliver the same benefits as CoMP-JT \cite{mobile2011c,Wu:2012:GreenWirelessComm}
but with less stringent synchronization requirements. In C-RANs, the
baseband (BB) signal processing units are no longer installed at base
stations (BSs) but relocated at a central cloud computing platform,
which is referred to as BB unit (BBU) pool. Thus, BSs on C-RANs are
solely responsible for wireless interface of the network, and now
called remote radio heads (RRHs). By these particular features, C-RANs
can potentially facilitate tight synchronization issue of BB signals
required for CoMP-JT technique, and also leverage powerful computing
capabilities for full cooperation \cite{Checko:2014:CRANRevew}. However,
BB signals from the BBU pool still need to be transported to the RRHs
through the fronthaul links of limited capacity. In addition, the
fronthaul links should support the strict latency and jitter requirements
in order to perform the synchronization across the connected RRHs.
Those are the main challenges of the C-RAN design in practice \cite{Checko:2014:CRANRevew,Peng:2015:FHInsightChallenge,Peng:ChanEst:Survey}.

Due to the growing concern over the power consumption in existing
mobile networks, recent research in wireless communications has shifted
its focus on energy efficiency (EE) approaches \cite{FengJiang:13:aSurveyofEE}.
In the past, wireless communications systems were mainly developed
to maximize the spectral efficiency, i.e., with the aim to transmit
at high data rates at any cost. This leads to a huge amount of power
consumption on current wireless networks, since high data rate transmission
essentially requires high transmit power. The notion of EE on the
other hand, measured in bits/Joule, considers the data rate and total
power consumption simultaneously.

C-RANs are a promising solution to address the problem of EE in future
wireless networks, i.e., 5G and beyond. The potential gains of C-RANs
on delivering the EE performance will be explored in this paper. Since
the RRHs are controlled by the common BBU pool, they can be switched
off to reduce the power consumption, thereby increasing the EE. In
doing so, we also need to assign users to a proper set of serving
RRHs. This should be done taking into account the limited capacity
of fronthaul links.

\subsection{Related Works}

The problem of EE maximization (EEmax) has been studied in prior publications
\cite{zappone2015energy,Dan:2013:SP:Fullduplex,DerrickKwanNg:2012:JWCOM:EE_OFDM,HHJY:13:JCOM,Giang:15:JCOML,Oskari:EE-optimalBeamDesign:14:JSP,zappone:2017:GlobalEE,HHJYL:13:JCOML}
for different contexts. In the noise-limited scenarios, \emph{parametric
fractional programming} (PFP), i.e.,\ Dinkelbach's algorithms were
used to globally solve the EE power control problems with a linear
(even super-linear) convergence \cite{zappone2015energy}. In multiuser
interference channels, Dinkelbach's algorithms cannot be applied to
the EEmax problems here since Dinkelbach's assumptions are not met
\cite{zappone2015energy}. Thus, \cite{Oskari:EE-optimalBeamDesign:14:JSP}
and \cite{zappone:2017:GlobalEE} resorted to using monotonic optimization
in order to achieve optimal EE solution for multiple-input single-output
(MISO) or single-input single-output (SISO) systems. As such global
optimization methods require prohibitively high complexity, efficient
suboptimal solutions were also of particular interest. Among them,
the heuristic approaches developed based on PFP and the \emph{successive
convex approximation} (SCA) have been widely used in many wireless
applications \cite{DerrickKwanNg:2012:JWCOM:EE_OFDM,HHJY:13:JCOM,Dan:2013:SP:Fullduplex,Giang:15:JCOML,Oskari:EE-optimalBeamDesign:14:JSP,zappone:2017:GlobalEE,HHJYL:13:JCOML}.
It is observed that the former approach often leads to a multi-stage
iterative procedure \cite{DerrickKwanNg:2012:JWCOM:EE_OFDM,HHJY:13:JCOM},
and the convergence may not be guaranteed \cite[Section 4.1]{zappone2015energy}.
On the other hand, the latter usually results in one-layer iterative
procedures provably converging to stationary points with a small number
of iterations \cite{Oskari:EE-optimalBeamDesign:14:JSP,Giang:15:JCOML}.
In fact, extensive numerical experiments conducted for some EEmax
problems in multiuser MISO systems showed that the SCA-based methods
outperform the heuristic PFP-based methods in terms of computational
complexity \cite{Giang:15:JCOML,Oskari:EE-optimalBeamDesign:14:JSP}.

The aforementioned and other related studies assume that signal processing
power is independent of the data rate. However, this is a simplification
as different data rates require different modulation and coding schemes.
In fact, signal processing power increases proportionally with the
data rates \cite{Isheden:2010:globecom,Xiong:2012:EEOFDMA,Wang:2013:EERate-dependentPower},\cite{Tran:2017:VirtualizedBBU}.
Moreover, the efficiency of the power amplifiers (PAs) is also assumed
to be a constant in previous studies on EEmax \cite{Dan:2013:SP:Fullduplex,DerrickKwanNg:2012:JWCOM:EE_OFDM,HHJY:13:JCOM,Giang:15:JCOML,Oskari:EE-optimalBeamDesign:14:JSP,zappone:2017:GlobalEE,HHJYL:13:JCOML,Hajisami:2017:ElasticNet}.
As shown in many works, PA's efficiency is often dynamic and it is
degraded when operating in the back-off region of the maximum power
\cite{Mikami2007,Bjoernemo2009,AmplifierMIMO-Persson}. Thus, it is
practically relevant to investigate the impact of the dynamic of PA's
efficiency and rate dependent power on EEmax designs.

C-RAN designs concerning limited fronthaul have been considered in
some recent works \cite{BinBinDai:15:Access:BeamformingBH,Dai2016:EECRAN,Tran:2017:DynamicCRAN,FuxinKLau:2014:SP:BackhaulLimitedSDPRelax,GiangICASSP2017}.
A simple, but effective and widely used, method is to reduce the amount
of BB signals exchanged through the fronthaul links. This is done
by selecting a set of users that can be served by a RRH, giving rise
to the RRH-user association problem that is often jointly designed
with the transmit beamforming to optimize a network performance measure
such as sum rate, power consumption or EE\textcolor{blue}{{} }\cite{BinBinDai:15:Access:BeamformingBH,Dai2016:EECRAN,Tran:2017:DynamicCRAN,FuxinKLau:2014:SP:BackhaulLimitedSDPRelax,GiangICASSP2017}.
In the multicarrier transmission, jointly optimizing RRH selection
and spectrum allocation would improve the network performances \cite{HAJISAMIDJP,HAJISAMICFFR}.
The RRH-user association and RRH selection problems are usually modeled
by a set of binary preference variables, leading to a \emph{mixed
binary integer program} (MBIP). As a result, optimal solutions to
C-RANs with RRH-user associations and RRH selection are difficult
to derive. On the other hand, in the EE perspective, there exists
an approach of minimizing total power consumption for improving EE
for C-RANs, e.g.,\  \cite{Dai2016:EECRAN,Hajisami:2017:ElasticNet}
for SISO or \cite{GroupGreenBeam-YShi} for MISO. However, since the
achieved data rate is not jointly considered in the objective, this
approach may be far from the optimal \cite{Oskari:EE-optimalBeamDesign:14:JSP,HHJYL:13:JCOML}.

\subsection{Contributions}

We investigate the EEmax problem in C-RANs with capacity-limited fronthaul.
Specifically, we propose a joint design of transmit beamforming and
RRH-user association to maximize the network EE, while satisfying
per-RRH fronthaul capacity, transmit power budget and users' quality-of-service
(QoS). Towards a more realistic power consumption model, we account
for the rate-dependent signal processing power and the dynamics of
PA's efficiency. The considered problem is modeled as an MBIP. Our
contributions include the following:
\begin{itemize}
\item We propose a globally optimal solution to the considered MBIP problem
by customizing the discrete branch-reduce-and-bound (DBRB) framework
introduced in \cite{tuy2006discrete}. To this end, we present transformations
to reformulate the design problem into a form that is amendable to
the application of the DBRB algorithm. Special modifications are made
to improve the convergence performance of the proposed method.
\item As global optimization methods are always of great concerns for an
MBIP, we also propose two suboptimal solutions to the joint design
problem that can achieve near-optimal solutions but with remarkably
reduced complexity. In the first method, we use a set of continuous
constraints to represent the binary variables, and then apply the
penalty method to solve the resulting problem. In the second one,
we approximate the binary variables by a piecewise linear function
which is inspired by \cite{le2015dc}. In both suboptimal methods,
the obtained continuous problems are nonconvex, which are solved by
the framework of the SCA.
\item We provide extensive numerical results to justify the proposed solutions.
The achievement of near-optimal performance by the proposed suboptimal
methods is demonstrated by benchmarking against the optimal one. We
compare the proposed solutions to other known methods in the literature.
The impacts of rate-dependent power and dynamic PA's efficiency are
also numerically investigated.
\end{itemize}
The rest of the paper is organized as follows. System model, design
constraints, power consumption model and problem formulation are described
in Section II. Section III presents the preliminaries of the DBRB
framework in solving an MBIP, followed by the customization to solve
the considered problem. Two suboptimal solutions are presented in
Section IV. Numerical results are provided in Section V and Section
VI concludes the paper.

\emph{Notation}: We follow the standard notations in this paper.
Lowercase letters, bold lowercase letters and bold uppercase letters
represent the scalars, column (row) vectors and matrices, respectively.
$\mathbb{Z}$, $\mathbb{R}$ and $\mathbb{C}$ represent the integer,
real and complex domains, respectively. $\left(.\right)^{T}$ and
$\left(.\right)^{H}$ represent the transpose and Hermitian transpose
operator, respectively. $\Re(.)$ and $|.|$ represent the real part
and absolute value of a complex number, respectively. $\|.\|_{2}$
represents the $\ell_{2}$ norm. The expectation of random variable
is denoted as $\mathbb{E}[\cdot]$. $\{\mathbf{a}_{b}\}_{b}$ and
$\{a_{b}\}_{b}$ refer to a set of vectors and scalars with different
index $b$, respectively. $[{\bf a}]_{i}$ is the $i$th element of
vector ${\bf a}$. ${\bf e}_{i}$ denotes the $i$th unit vector,
i.e., the vector such that $e_{i}=1$, $e_{j}=0$ $\forall j\neq i$.
 Finally, $\left\lceil a\right\rceil _{{\cal S}}$ and $\left\lfloor a\right\rfloor _{{\cal S}}$
are the upper and lower nearest neighbor elements of $a$ in set ${\cal S}$.

\section{System Model and Problem Formulation}

\subsection{System Model}

We consider a multiuser MISO wireless system consisting of a set of
$B$ RRHs, denoted by ${\cal B}\triangleq\{1,\ldots,B\}$, each equipped
with $I$ antennas\footnote{Herein, the same number of equipped antennas for all RRHs is assumed
purely for notational simplicity.}, and a set of $K$ single-antenna users, denoted by ${\cal K}\triangleq\{1,\ldots,K\}$.
The RRHs are connected to a common BBU pool through finite-capacity
fronthaul links. The BBU pool is assumed to achieve perfect channel
state information (CSI) associated with all the users in the network\textcolor{blue}{.}\footnote{From a practical perspective, overhead and accuracy of channel estimation
should be considered, since they have major impacts on the scale of
coordination and performance of C-RANs (see detail discussion in \cite{Peng:2015:FHInsightChallenge,Peng:ChanEst:Survey,Tran:2017:DynamicCRAN}).
Also, there are some channel estimation techniques proposed for C-RANs
which are summarized in \cite[Section V]{Peng:ChanEst:Survey}.} In this paper, CoMP-JT is considered, i.e., any user can simultaneously
receive data from multiple RRHs \cite{marsch2011coordinated}. Let
$d_{k}$ denote the data symbol intended for user $k$ which has unit-energy,
i.e., $\mathbb{E}[|d_{k}|^{2}]=1$, and ${\bf w}_{b,k}\in\mathbb{C}^{I\times1}$
denote the beamforming vector from RRH $b$ to user $k$. Assuming
a flat fading channel model, the received signal at user $k$ can
be written as
\begin{equation}
y_{k}=\underset{\textrm{desired\,signal}}{\underbrace{\left(\sum_{b\in{\cal B}}{\bf h}_{b,k}{\bf w}_{b,k}\right)d_{k}}}+\underset{\textrm{interference}}{\underbrace{\sum_{j\in{\cal K}\setminus k}\left(\sum_{b\in{\cal B}}\mathbf{h}_{b,k}\mathbf{w}_{b,j}\right)d_{j}}}+n_{k}
\end{equation}
where ${\bf h}_{b,k}\in\mathbb{C}^{1\times I}$ is the channel between
RRH $b$ and user $k$, and $n_{k}\sim\mathcal{CN}(0,\sigma_{k}^{2})$
is the additive white Gaussian noise at user $k$. For notational
convenience, let ${\bf h}_{k}\triangleq[{\bf h}_{1,k},{\bf h}_{2,k},\ldots,{\bf h}_{B,k}]\in\mathbb{C}^{1\times IB}$
and ${\bf w}_{k}\triangleq[{\bf w}_{1,k}^{T},{\bf w}_{2,k}^{T},\ldots,{\bf w}_{B,k}^{T}]^{T}$
$\in\mathbb{C}^{IB\times1}$ be the aggregate vectors of all channels
and beamformers from all RRHs to user $k$, respectively. We also
denote by ${\bf w}$ the beamforming vector stacking all ${\bf w}_{k}$.
Assuming single-user decoding, i.e. interference among users is treated
as Gaussian noise, the SINR at user $k$ can be written as
\begin{align}
\gamma_{k}({\bf w})\triangleq & \frac{|\sum_{b\in{\cal B}}{\bf h}_{b,k}{\bf w}_{b,k}|^{2}}{\sum_{j\in{\cal K}\setminus k}|\sum_{b\in{\cal B}}\mathbf{h}_{b,k}\mathbf{w}_{b,j}|^{2}+\sigma_{k}^{2}}\nonumber \\
= & \frac{|{\bf h}_{k}{\bf w}_{k}|^{2}}{{\textstyle \sum_{j\in{\cal K}\setminus k}}|{\bf h}_{k}{\bf w}_{j}|^{2}+\sigma_{k}^{2}}.
\end{align}
Let $r_{k}$ be the achievable data rate transmitted to user $k$.
By the Shannon's coding theory, we have
\[
r_{k}\leq\log(1+\gamma_{k}({\bf w})).
\]

\subsection{Fronthaul Constraints}

In practice the fronthaul link from the BBU pool to RRH $b$ has a
finite capacity, denoted by $\bar{C}_{b}$. To be feasible, the total
data rate of the wireless physical layer of RRH $b$ should not be
larger than $\bar{C}_{b}$. For the problem formulation purposes,
let us define $x_{b,k}\in\{0,1\}$ to be the preference variable representing
the connection between RRH \textbf{$b$ }and user $k$, i.e., $x_{b,k}=1$
indicates that user $k$ receives data from RRH $b$ and $x_{b,k}=0$
otherwise. Then it is clear that the total data rate which can be
reliably transmitted by the wireless interface of RRH $b$ is ${\textstyle \sum_{k\in{\cal K}}}x_{b,k}r_{k}$,
and thus the following constraint
\[
\sum_{k\in{\cal K}}x_{b,k}r_{k}\leq\bar{C}_{b}
\]
 should hold for RRH $b$.

\subsection{Power Consumption Model}

We consider the power consumption model based on those in \cite{Isheden:2010:globecom,HowmuchEnergy-Auer,Dhaini:2014:EE-TDMA,AmplifierMIMO-Persson}
which includes the power consumed by the electronic circuits in the
network and the PAs on RRHs. Specifically, the circuit power consumption
is divided into two parts as detailed below.

\subsubsection{Rate-independent Circuit Power Consumption}

The rate-independent power consumption is modeled as \cite{HowmuchEnergy-Auer,Dhaini:2014:EE-TDMA}
\begin{equation}
\begin{alignedat}{1}P_{\text{I}}\triangleq\  & KP_{\text{ms}}+\sum_{b\in{\cal B}}s_{b}\underset{\textrm{active\,mode}}{\underbrace{(P_{\text{RRH}}^{\textrm{active}}+P_{\text{NU}}^{\textrm{active}})}}\\
 & +\sum_{b\in{\cal B}}(1-s_{b})\underset{\textrm{sleep\,mode}}{\underbrace{(P_{\text{RRH}}^{\textrm{sleep}}+P_{\text{NU}}^{\textrm{sleep}})}}+P_{\text{OLT}}.
\end{alignedat}
\label{eq:inde-powermodel}
\end{equation}
In \eqref{eq:inde-powermodel}, $P_{\text{ms}}$ is the circuit power
consumed by a user device, $P_{\text{RRH}}^{\textrm{active}}$ and
$P_{\text{RRH}}^{\textrm{sleep}}$ are the power consumption at a
RRH corresponding to the active and sleep modes, respectively. In
particular, $P_{\text{RRH}}^{\textrm{active}}$ consists of power
for feeding signal processing circuits of transceiver chains, and
operating RRHs (e.g. main supply, site-cooling) and hardware elements
for RF parts (e.g. converters, filters, mixers, etc) \cite{HowmuchEnergy-Auer}.
It is assumed that all RRHs connect to the BBU pool through a passive
optical network which consists of an optical line terminal (OLT) and
a set of network units (NUs) \cite{Dhaini:2014:EE-TDMA}. The OLT
is always active and consumes a fixed power, i.e. $P_{\text{OLT}}$
in \eqref{eq:inde-powermodel}. On the other hand, NUs are switchable
between the active and sleep modes for power saving purposes, each
consuming a power $P_{\text{NU}}^{\textrm{active}}$ and $P_{\text{NU}}^{\textrm{sleep}}$,
respectively. In order to represent the operating mode of RRH $b$
and the associated NU, we introduce binary preference variables $\{s_{b}\}_{b}$
such that $s_{b}=1$ when RRH and NU $b$ is active and $s_{b}=0$
otherwise. The relationship between $s_{b}$ and $x_{b,k}$ (introduced
in the previous subsection) can be represented as
\begin{equation}
s_{b}=\max_{k\in{\cal K}}\{x_{b,k}\}\Leftrightarrow\begin{cases}
s_{b}\geq x_{b,k},\forall k\in{\cal K}\\
s_{b}\leq\sum_{k\in{\cal K}}x_{b,k}
\end{cases},\forall b\in{\cal B}
\end{equation}
i.e., $s_{b}=1$ when RRH $b$ serves at least one user and $s_{b}=0$
otherwise.

\subsubsection{Rate-dependent BB Signal Processing Power}

The power consumed by the signal processing operations at the BBU
pool such as channel encoding, decoding and fronthauling expenditure
depends on the data rate \cite{Isheden:2010:globecom,Xiong:2012:EEOFDMA,Wang:2013:EERate-dependentPower,Tran:2017:VirtualizedBBU}.
For RRH $b$, this power consumption is measured by a continuous function
of the fronthaul rate $\tilde{r}_{b}$ denoted as $\psi_{b}(\tilde{r}_{b})$
where $\tilde{r}_{b}\triangleq{\textstyle \sum_{k\in{\cal K}}}x_{b,k}r_{k}$.
According to \cite{Isheden:2010:globecom}\textcolor{blue}{,}\cite{Tran:2017:VirtualizedBBU},
$\psi_{b}(\tilde{r}_{b})$ is linearly scaled w.r.t. $\tilde{r}_{b}$,
i.e.,
\begin{equation}
\psi_{b}(\tilde{r}_{b})=p_{\text{SP}}\tilde{r}_{b}
\end{equation}
where $p_{\text{SP}}$ is a constant coefficient in $\text{W/(Gnats/s)}$.

\subsubsection{Dynamic Power Amplifier}

Many existing approaches in relation to energy-efficient design assume
a constant efficiency of PAs in their problem formulation \cite{Dan:2013:SP:Fullduplex,DerrickKwanNg:2012:JWCOM:EE_OFDM,HHJY:13:JCOM,Giang:15:JCOML,zappone:2017:GlobalEE,Oskari:EE-optimalBeamDesign:14:JSP}.
However, in practice, the efficiency of PAs depends on their operating
conditions, and thus is dynamic \cite{Mikami2007,Bjoernemo2009,AmplifierMIMO-Persson}.
We can model the PA's efficiency of RF chain $i$ at RRH $b$ as \cite{AmplifierMIMO-Persson}
\begin{equation}
\epsilon_{b,i}(\{{\bf w}_{b,k}\}_{k})\triangleq\frac{1}{\tilde{\epsilon}}\sqrt{\sum_{k\in{\cal K}}|[\mathbf{w}_{b,k}]_{i}|^{2}}\label{eq:nonlinearpoweff}
\end{equation}
 where $\tilde{\epsilon}\triangleq\sqrt{P_{\textrm{a}}}/\epsilon_{\max}$,
and $P_{\textrm{a}}$ and $\epsilon_{\max}\in[0,1]$ are the maximum
power of the PA and the maximum PA's efficiency, respectively. Let
$\phi_{b}(\{{\bf w}_{b,k}\}_{k})$ be a function of beamforming vectors
which measures the amount of power consumed by the PAs for radiating
the transmitted signals outwards the antennas at RRH $b$. From \eqref{eq:nonlinearpoweff},
$\phi_{b}(\{{\bf w}_{b,k}\}_{k})$ is expressed as
\begin{equation}
\phi_{b}(\{{\bf w}_{b,k}\}_{k})=\sum_{i=1}^{I}\frac{\sum_{k\in{\cal K}}|[\mathbf{w}_{b,k}]_{i}|^{2}}{\epsilon_{b,i}(\{{\bf w}_{b,k}\}_{k})}=\tilde{\epsilon}\sum_{i=1}^{I}||\tilde{\mathbf{w}}_{b,i}||_{2}
\end{equation}
where $\tilde{\mathbf{w}}_{b,i}\triangleq[[\mathbf{w}_{b,1}]_{i};[\mathbf{w}_{b,2}]_{i};...;[\mathbf{w}_{b,K}]_{i}]\in\mathbb{C}^{K\times1}$.

\subsubsection{Total Power Consumption}

For notational convenience, let us define ${\bf x}\triangleq\{x_{b,k}\}_{b\in{\cal B},\!k\in{\cal K}}$,
${\bf s}\triangleq\{s_{b}\}_{b\in{\cal B}}$, and ${\bf r}\triangleq\{r_{k}\}_{k\in{\cal K}}$.
Based on the above discussions, the total consumed power in the considered
system is denoted by $f_{\textrm{P}}({\bf w},{\bf x},{\bf r},{\bf s})$
and can be expressed as
\begin{equation}
\begin{alignedat}{1} & f_{\textrm{P}}({\bf w},{\bf x},{\bf r},{\bf s})\triangleq P_{\text{I}}+\sum_{b\in{\cal B}}\left(\psi_{b}(\tilde{r}_{b})+\phi_{b}(\{{\bf w}_{b,k}\}_{k})\right)\\
 & =\sum_{b\in{\cal B}}\Bigl(\tilde{\epsilon}\sum_{i=1}^{I}||\tilde{\mathbf{w}}_{b,i}||_{2}+\Delta Ps_{b}+p_{\text{SP}}\sum_{k\in{\cal K}}x_{b,k}r_{k}\Bigr)\\
 & \qquad\qquad\qquad+\underbrace{BP^{\textrm{sleep}}+KP_{\text{ms}}+P_{\text{OLT}}}_{P_{\text{const}}}
\end{alignedat}
\label{eq:power model}
\end{equation}
in which $P^{\textrm{active}}\triangleq P_{\text{RRH}}^{\textrm{active}}+P_{\text{NU}}^{\textrm{active}}$,
$P^{\textrm{sleep}}\triangleq P_{\text{RRH}}^{\textrm{sleep}}+P_{\text{NU}}^{\textrm{sleep}}$
and $\Delta P\triangleq P^{\textrm{active}}-P^{\textrm{sleep}}$ which
are constants.

\subsection{Problem Formulation}

We consider the problem of joint beamforming and RRH-user association
design where the overall network EE is maximized. Mathematically,
the problem of interest reads \begin{subequations}\label{Prob:Gen:Problem}
\begin{align}
\underset{{\bf w},{\bf x},{\bf s},{\bf r}}{\text{maximize}}\  & \ \frac{\sum_{k\in{\cal K}}r_{k}}{f_{\textrm{P}}({\bf w},{\bf x},{\bf r},{\bf s})}\label{eq:obj:gen:prob}\\
\text{subject to}\  & \ r_{k}\leq\log(1+\gamma_{k}({\bf w})),\ \forall k\in{\cal K}\label{eq:achievable rate}\\
 & \ r_{k}\geq r_{0},\ \forall k\in{\cal K}\label{eq:QoS:constraint}\\
 & \sum_{k\in{\cal K}}x_{b,k}r_{k}\leq\bar{C}_{b},\ \forall b\in{\cal B}\label{eq:BH:constraint}\\
 & \sum_{k\in{\cal K}}\|{\bf w}_{b,k}\|_{2}^{2}\leq\bar{P}_{b},\ \forall b\in{\cal B}\label{eq:BS:powerconstraint}\\
 & ||\tilde{\mathbf{w}}_{b,i}||_{2}^{2}\leq P_{\textrm{a}},\ \forall b\in{\cal B},i=1,...,I\label{eq:per-antenna:powerconstraint}\\
 & \|{\bf w}_{b,k}\|_{2}^{2}\leq x_{b,k}\bar{P}_{b},\ \forall k\in{\cal K},b\in{\cal B}\label{eq:beam:powerconstraint}\\
 & \sum_{b\in{\cal B}}x_{b,k}\geq1,\forall k\in{\cal K}\label{eq:min:connection}\\
 & s_{b}\geq x_{b,k},\forall k\in{\cal K};\ s_{b}\leq\sum_{k\in{\cal K}}x_{b,k},\ \forall b\in{\cal B}\label{eq:selection:relation}\\
 & {\bf x}\in\{0,1\}^{BK},\ {\bf s}\in\{0,1\}^{B}.\label{eq:Boolean:constraint}
\end{align}
\end{subequations}We impose \eqref{eq:QoS:constraint} to guarantee
that the data rate of user $k$ is not smaller than $r_{0}$ to meet
the required QoS. The constraints \eqref{eq:BS:powerconstraint} and
\eqref{eq:per-antenna:powerconstraint} represent the total transmit
power and per antenna power constraints at each individual RRH, respectively.
The constraints in \eqref{eq:beam:powerconstraint} guarantee that
if RRH $b$ does not serve user $k$, i.e. $x_{b,k}=0$ then it holds
that $\|{\bf w}_{b,k}\|_{2}^{2}=0$. The constraints in \eqref{eq:min:connection}
imply that each user is served by at least one RRH (due to the required
QoS).

We remark that Dinkelbach's algorithm cannot be applied to find optimal
solutions of \eqref{Prob:Gen:Problem}, since \eqref{eq:obj:gen:prob}
is intractable \cite[Section 3]{zappone2015energy}.\footnote{Applying Dinkelbach's method to (9) results in the parametric subproblem
(solved in each iteration) which is still nonconvex.} In fact, problem \eqref{Prob:Gen:Problem} is a nonconvex MBIP generally
known to be NP-hard\textcolor{blue}{. }In the following sections we
first derive an optimal algorithm to solve \eqref{Prob:Gen:Problem}
by customizing the DBRB framework, and then propose low-complexity
suboptimal approaches that can achieve the near-optimal performance.

\section{Optimal Jointly Energy-efficient Beamforming and RRH-user Association
Design}

\emph{General monotonic optimization} (GMO) is a widely-used global
continuous optimization technique \cite{tuy2005monotonic} for solving
numerous wireless communications nonconvex problems \cite{Bjornson2012RobustMonotonicMISO,Jorswieck2010MonotonicMISO,zappone:2017:GlobalEE,Oskari:EE-optimalBeamDesign:14:JSP}.
For MBIP problems, the GMO principle is inapplicable, since it outputs
only approximate solutions of discrete variables at convergence \cite{tuy2005monotonic}.
In recent work of \cite{PhuongTSP2017}, Luong \emph{et al.} combined
GMO with mixed integer programming (MIP) to solve their considered
problem which is also an MBIP. Particularly, the GMO works on the
continuous domain of their problem, and at each iteration of GMO,
a mixed integer program is solved. In this paper, we propose below
a new globally optimal approach to solve \eqref{Prob:Gen:Problem}
based on the so-called \emph{discrete} \emph{monotonic optimization}
(DMO) \cite{tuy2006discrete}.

\subsection{Preliminaries: Discrete Branch-reduce-and-bound}

To proceed we provide some background of DMO and briefly review the
DBRB procedure. In this paper we follow the definitions of \emph{box},
\emph{increasing function}, and \emph{normal cone} in \cite{tuy2006discrete}.
The standard form of a DMO problem is given by \cite{tuy2006discrete}
\begin{equation}
\max_{{\bf y}}\ f({\bf y})\ \text{subject to}\ \{{\bf y}\in{\cal S}\subseteq D\triangleq[{\bf a};{\bf b}]\}\label{eq:Gen:prob}
\end{equation}
where $f({\bf y})$ is an increasing function w.r.t. variable ${\bf y}$;
${\bf y}\triangleq[{\bf y}_{\text{d}}^{T},{\bf y}_{\text{c}}^{T}]^{T}\in\mathbb{R}^{N_{\text{d}}+N_{\text{c}}}$,
${\bf y}_{\text{d}}\in\mathbb{Z}^{N_{\text{d}}}$ and ${\bf y}_{\text{c}}\in\mathbb{R}^{N_{\text{c}}}$
are the discrete and continuous variables respectively; ${\cal S}$
is normal feasible set of ${\bf y}$; and $D$ is the box containing
${\cal S}$ with lower and upper vertices ${\bf a}$ and ${\bf b}$,
respectively.

\subsubsection{DBRB Procedure}

Similar to the standard branch-reduce-and-bound (BRB) algorithm \cite{tuy2005monotonic},
DBRB is an iterative procedure performing three basic operations at
each iteration:\emph{ branching}, \emph{reduction, }and \emph{bounding}.
Starting from original box $[{\bf a};{\bf b}]$, we iteratively divide
it into smaller and smaller ones, remove boxes that do not contain
an optimal solution, search over remaining boxes for an improved solution
until an error tolerance is met. Since the feasible set of the discrete
optimization problem is smaller than that of its continuous relaxation,
DBRB is modified from the standard BRB procedure in order to efficiently
remove those regions not belonging to the discrete constraints, thereby
achieving exact solutions \cite{tuy2006discrete}. In particular,
during the branching and reduction steps, elements corresponding to
discrete constraints are adjusted to stay in the discrete set. Details
of these three operations are presented next.

\subsubsection*{Branching}

\begin{figure}
\centering{}\includegraphics[width=1\columnwidth]{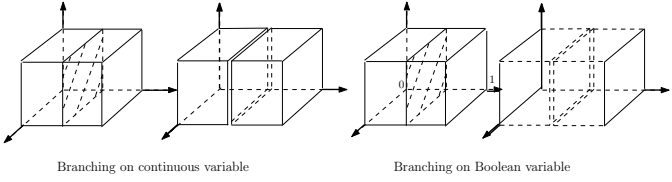}\caption{Illustration for branching operator.}
\label{branching:illustration}
\end{figure}

At iteration $n$ we select a box in the set of candidate boxes, denoted
by ${\cal R}_{n}$, and split it into two new boxes, which are of
equal size. To be bound improving we pick a box $V_{\text{c}}\triangleq[{\bf p};{\bf q}]\in{\cal R}_{n}$,
which has the largest upper bound, i.e., $V_{\text{c}}=\arg\max_{V\in{\cal R}_{n}}f_{\text{U}}(V)$
($f_{\text{U}}(V)$ denotes the upper bound of $V$), and bisect along
the longest edge, i.e., $l=\arg\max_{1\leq j\leq N_{\text{d}}+N_{\text{c}}}(q_{j}-p_{j})$
to create two smaller boxes $V_{\text{c}}^{1}=[{\bf p};{\bf q}^{\prime}]$
and $V_{\text{c}}^{2}=[{\bf p}^{\prime};{\bf q}]$, in which ${\bf q}^{\prime}$
and ${\bf p}^{\prime}$ are given\,by
\begin{equation}
q_{j}^{\prime}=\begin{alignedat}{1}\begin{cases}
q_{j} & \forall j\neq l\\
\left\lfloor q_{j}-(q_{j}-p_{j})/2\right\rfloor _{\mathbb{Z}} & \bm{\text{if }}j=l\leq N_{\text{d}},\\
q_{j}-(q_{j}-p_{j})/2 & \bm{\text{if }}j=l>N_{\text{d}},
\end{cases}\end{alignedat}
\end{equation}
and
\begin{equation}
p_{j}^{\prime}=\begin{cases}
p_{j} & \forall j\neq l\\
\left\lceil p_{j}+(q_{j}-p_{j})/2\right\rceil _{\mathbb{Z}} & \bm{\text{if }}j=l\leq N_{\text{d}},\\
p_{j}+(q_{j}-p_{j})/2 & \bm{\text{if }}j=l>N_{\text{d}},
\end{cases}
\end{equation}
respectively.
\begin{rem}
\label{rem:Boolean:branch}(\emph{Branching over Binary variable}s)
If $p_{j},q_{j}\in\{0,1\}$ and $q_{j}-p_{j}=1$ for $j\leq N_{\text{d}}$,
then $\left\lfloor q_{j}-(q_{j}-p_{j})/2\right\rfloor _{\{0,1\}}=0$
and $\left\lceil p_{j}+(q_{j}-p_{j})/2\right\rceil _{\{0,1\}}=1$
(e.g. see Fig. \ref{branching:illustration}).
\end{rem}

\subsubsection*{Reduction}

For any box, it possibly contains segments either infeasible to \eqref{eq:Gen:prob}
or resulting in an objective smaller than the \emph{current best objective}
(CBO), i.e. the known feasible point that offers the best objective
value at current iteration. Reduction is to remove those portions
of no interest to reduce the search space in the next iterations.
\begin{figure}
\centering{}\includegraphics[width=1\columnwidth]{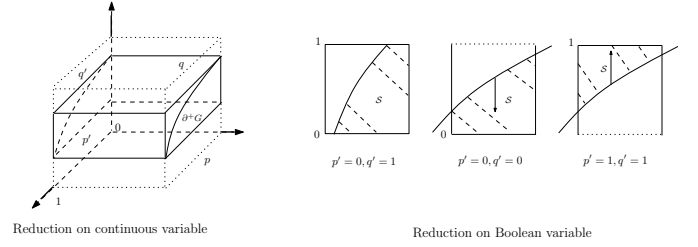}\caption{Illustration for reduction operator.}
\label{reduction:illustration}
\end{figure}
 Given a box $V=[{\bf p};{\bf q}]$, we wish to shrink the size of
$V$ without loss of optimality by creating a smaller box $\re(V)\triangleq[{\bf p}^{\prime};{\bf q}^{\prime}]\subset V$
such that an optimal solution (if exists in $V$) must be contained
in $\re(V)$. To do so we eliminate the portions $[{\bf p};{\bf p}^{\prime})$
and $({\bf q}^{\prime};{\bf q}]$ that result in an objective value
smaller than the CBO and/or are infeasible to \eqref{eq:Gen:prob}.
Mathematically, we can replace ${\bf p}$ by ${\bf p}^{\prime}\geq{\bf p}$
where ${\bf p}^{\prime}={\bf q}-{\textstyle \sum_{j=1}^{N_{\text{d}}+N_{\text{c}}}}\alpha_{j}(q_{j}-p_{j}){\bf e}_{j}$
and
\begin{equation}
\begin{alignedat}{1}\alpha_{j}=\sup\{\alpha\ |0\leq\alpha\leq1, & \ {\bf q}-\alpha(q_{j}-p_{j}){\bf e}_{j}\in D\backslash{\cal S},\\
 & f({\bf q}-\alpha(q_{j}-p_{j}){\bf e}_{j})\geq\text{CBO}\}
\end{alignedat}
\label{eq:find alpha}
\end{equation}
 for each $j=1,\ldots,N_{\text{d}}+N_{\text{c}}$. Similarly, vertex
${\bf q}$ is replaced by ${\bf q}^{\prime}\leq{\bf q}$ where ${\bf q}^{\prime}={\bf p}^{\prime}+{\textstyle \sum_{j=1}^{N_{\text{d}}+N_{\text{c}}}}\beta_{j}(q_{j}-p_{j}^{\prime}){\bf e}_{j}$
and
\begin{equation}
\begin{alignedat}{1}\beta_{j}=\sup\{\beta\ | & 0\leq\beta\leq1,\ {\bf p}^{\prime}+\beta(q_{j}-p_{j}^{\prime}){\bf e}_{j}\in{\cal S}\}.\end{alignedat}
\label{eq:find beta}
\end{equation}
The values of $\alpha_{j}$ and $\beta_{j}$ in \eqref{eq:find alpha}
and \eqref{eq:find beta} can be found easily by the bisection method.
Note that for $j\leq N_{\text{d}}$, the output of the reduction procedure
is then adjusted into the discrete set, i.e., $p_{j}^{\prime}=\left\lceil p_{j}^{\prime}\right\rceil _{\mathbb{Z}}$
and $q_{j}^{\prime}=\left\lfloor q_{j}^{\prime}\right\rfloor _{\mathbb{Z}}$.
\begin{rem}
\label{rem:Boolean:reduce}(\emph{Reduction over Binary variable}s)
If $p_{j},q_{j}\in\{0,1\}$ and $q_{j}-p_{j}=1$ for $j\leq N_{\text{d}}$,
we can quickly set that $p_{j}^{\prime}=\begin{cases}
1 & \text{if }\ {\bf q}-{\bf e}_{j}\in D\backslash{\cal S}\\
0 & \text{otherwise},
\end{cases}$. If $p_{j}^{\prime}=0$, we then replace $q_{j}-p_{j}^{\prime}=1$
into \eqref{eq:find beta} and obtain $q_{j}^{\prime}=\begin{cases}
1 & \text{if }\ {\bf p}^{\prime}+{\bf e}_{j}\in{\cal S}\\
0 & \text{otherwise}
\end{cases}$ (e.g. see Fig. \ref{reduction:illustration}).
\end{rem}
The reduction procedure above does not drop off any feasible solution
of \eqref{eq:Gen:prob} as shown in \cite{tuy2006discrete}.

\subsubsection*{Bounding}

Bounding is another basic operation for the DBRB to ensure the convergence.
The main purpose of this step is to improve the upper and lower bounds
of $f({\bf y})$. Due to its monotonicity, the upper and lower bounds
of a box $V=[\mathbf{p};\mathbf{q}]$ can be easily found as $f(\mathbf{p})$
and $f(\mathbf{q})$, respectively. These bounds are then used to
update the CBO as mentioned above and to remove the boxes whose upper
bound is smaller than the CBO \cite{tuy2006discrete}.

We are now ready to customize the DBRB procedure to solve problem
\eqref{Prob:Gen:Problem}. Algorithm~\ref{Alg. BRB} outlines our
proposed optimal method and its details are presented in the sequel.
\begin{algorithm}[t]
\caption{The proposed DBRB algorithm }
\label{Alg. BRB}

\begin{algorithmic}[1]

\STATE \textbf{Initialization:} Compute ${\bf a}$, ${\bf b}$ and
apply box reduction to box $[{\bf a};{\bf b}]$. Let $n:=1$, ${\cal R}_{1}=\re([{\bf a};{\bf b}])$
and $\eta_{1}^{\text{best}}=0$

\REPEAT[\textbf{$n:=n+1$}.]

\STATE{\textbf{Branching:} select a box $V_{\text{c}}=[{\bf p};{\bf q}]\subset{\cal R}_{n-1}$
and branch $V_{\text{c}}$ into two smaller ones $V_{\text{c}}^{1}$
and $V_{\text{c}}^{2}$, then remove $V_{\text{c}}$ from ${\cal R}_{n-1}$.}

\STATE{\textbf{Reduction}: apply box reduction to each box $V_{\text{c}}^{m}$
($m=\{1,2\})$ and obtain reduced box $\re(V_{\text{c}}^{m})$.}

\STATE{\textbf{Bounding}: for each box $\re(V_{\text{c}}^{m})$ not
violating \eqref{eq:bh:feasibility}}\label{Bounding}

\IF{solving \eqref{Prob:SOCP:checkfeasible} is feasible}

\STATE{Achieve ${\bf w}^{\ast},{\bf u}^{\ast}$, calculate ${\bf t}^{\ast}$
and extract ${\bf x}^{\ast}$.}

\STATE{Update $\underline{{\bf t}}:={\bf t}^{\ast}$ and calculate
$\eta_{\text{U}}(\re(V_{\text{c}}^{m}))$ by \eqref{eq:eta:upbound}.}

\STATE{Check ${\bf x}^{\ast}$ with \eqref{eq:feasible:cond}, if
true, obtain $\eta_{\text{L}}(\re(V_{\text{c}}^{m}))$ as \eqref{eq:eta:lowbound}
and update CBO $\eta_{n}^{\text{best}}:=\max\{\eta_{\text{L}}(\re(V_{\text{c}}^{m})),\eta_{n-1}^{\text{best}}\}$,
otherwise $\eta_{\text{L}}(\re(V_{\text{c}}^{m}))=\frac{\sum_{k\in{\cal K}}\underline{r}_{k}}{\hat{f}_{\text{P}}(\overline{{\bf s}},\overline{{\bf x}},\overline{{\bf r}},\overline{{\bf t}})}$
.}\label{search}

\STATE{Update ${\cal R}_{n}:={\cal R}_{n-1}\cup\{\re(V_{\text{c}}^{m})|\eta_{\text{U}}(\re(V_{\text{c}}^{m}))\geq\eta_{n}^{\text{best}}\}$.}\label{remove}

\ENDIF

\UNTIL{Convergence}

\STATE \textbf{Output:} With $(\eta_{n}^{\text{best}},{\bf x}^{\ast},{\bf s}^{\ast},{\bf r}^{\ast},{\bf t}^{\ast})$,
recover ${\bf w}^{\ast}$ by \eqref{eq:find beam} to achieve the
globally optimal solution of \eqref{Prob:Gen:Problem}, i.e.\ $({\bf w}^{\ast},{\bf x}^{\ast},{\bf s}^{\ast},{\bf r}^{\ast})$.

\end{algorithmic}
\end{algorithm}

\subsection{Customization of DBRB for Solving \eqref{Prob:Gen:Problem}\label{subsec:Customization-of-DBRB}}

We remark that \eqref{Prob:Gen:Problem} is not a DMO problem in a
standard form, since the objective in \eqref{eq:obj:gen:prob} is
not an increasing function w.r.t. the involved variables. To apply
the DBRB algorithm we first reformulate \eqref{Prob:Gen:Problem}
as \begin{subequations}\label{Prob:EE:epi}
\begin{align}
\underset{\eta,\substack{{\bf w},{\bf x},{\bf s},{\bf r},{\bf t}}
}{\text{maximize}} & \quad\eta\\
\text{subject to} & \quad\eta\hat{f}_{\text{P}}({\bf x},{\bf s},{\bf r},{\bf t})-{\textstyle \sum_{k\in{\cal K}}}r_{k}\leq0\label{eq:EE:epi:constraint}\\
 & \quad{\textstyle \sum_{i=1}^{I}}||\tilde{\mathbf{w}}_{b,i}||_{2}\leq t_{b},\ \forall b\in{\cal B}\label{eq:antenna:powerconsumption}\\
 & \quad\eqref{eq:achievable rate}-\eqref{eq:Boolean:constraint}
\end{align}
\end{subequations} where $\eta$ and ${\bf t}\triangleq\{t_{b}\}_{b}$
are newly introduced variables and $f_{\text{P}}({\bf w},{\bf x},{\bf s},{\bf r})$
is redefined as $\hat{f}_{\text{P}}({\bf x},{\bf s},{\bf r},{\bf t})\triangleq\sum_{b\in{\cal B}}\left(\tilde{\epsilon}t_{b}+\Delta Ps_{b}+p_{\text{SP}}\sum_{k\in{\cal K}}x_{b,k}r_{k}\right)+P_{\text{const}}$.
The equivalence between \eqref{Prob:Gen:Problem} and \eqref{Prob:EE:epi}
in terms of optimal solution set can be easily proved, since \eqref{Prob:EE:epi}
is indeed the epigraph of \eqref{Prob:Gen:Problem}. Towards solving
\eqref{Prob:EE:epi} we have the following lemma.
\begin{lem}
\label{lem:1}Let $(\eta^{\ast},{\bf w}^{\ast},{\bf x}^{\ast},{\bf s}^{\ast},{\bf r}^{\ast},{\bf t}^{\ast})$
denote an optimal solution to \eqref{Prob:EE:epi}. Given the value
of $({\bf x}^{\ast},{\bf s}^{\ast},{\bf r}^{\ast},{\bf t}^{\ast})$,
then the optimal beamforming vector, denoted by ${\bf w}^{\ast}$,
can be computed as
\begin{equation}
{\bf w}^{\ast}=\mathrm{find}\{{\bf w}|\eqref{eq:achievable rate},\eqref{eq:BS:powerconstraint}-\eqref{eq:beam:powerconstraint},\eqref{eq:antenna:powerconsumption}\}\label{eq:find beam}
\end{equation}
in which we replace $({\bf x},{\bf s},{\bf r},{\bf t})$ by $({\bf x}^{\ast},{\bf s}^{\ast},{\bf r}^{\ast},{\bf t}^{\ast})$.
\end{lem}
\begin{IEEEproof}
See Appendix A.
\end{IEEEproof}
The lemma implies that we can obtain ${\bf w}^{\ast}$ if $({\bf x}^{\ast},{\bf s}^{\ast},{\bf r}^{\ast},{\bf t}^{\ast})$
are known. We remark that $\eta$ is easily determined when $({\bf x},{\bf s},{\bf r},{\bf t})$
is fixed as $\eta=\frac{\sum_{k\in{\cal K}}r_{k}}{\hat{f}_{\text{P}}({\bf x},{\bf s},{\bf r},{\bf t})}$.
Also, the feasibility of ${\bf r}$ depends on ${\bf t}$, ${\bf x}$
and ${\bf s}$ as can be seen in \eqref{eq:BH:constraint} and \eqref{eq:EE:epi:constraint}.
Furthermore constraints \eqref{eq:BH:constraint}, \eqref{eq:min:connection},
\eqref{eq:selection:relation} and \eqref{eq:EE:epi:constraint} are
monotone w.r.t. ${\bf x}$, ${\bf s}$, ${\bf r}$ and ${\bf t}$.
Thus we can develop a DBRB algorithm to solve \eqref{Prob:EE:epi}
by branching over $({\bf x},{\bf s},{\bf r},{\bf t})$, which is the
central idea of the proposed algorithm as described next.

Let ${\cal S}$ be the feasible set of problem \eqref{Prob:EE:epi},
i.e.,
\[
\begin{alignedat}{1}{\cal S}\triangleq & \{[{\bf x},{\bf s},{\bf r},{\bf t}]|\eqref{eq:achievable rate},\eqref{eq:QoS:constraint},\eqref{eq:per-antenna:powerconstraint}-\eqref{eq:Boolean:constraint},\eqref{eq:EE:epi:constraint},\\
 & \qquad\qquad\sum_{k\in{\cal K}}x_{b,k}r_{k}\leq s_{b}\bar{C}_{b},\sum_{k\in{\cal K}}\|{\bf w}_{b,k}\|_{2}^{2}\leq s_{b}\bar{P}_{b},\\
 & \qquad\qquad\sum_{i=1}^{I}||\tilde{\mathbf{w}}_{b,i}||_{2}\leq s_{b}t_{b},\forall b\in{\cal B}\}.
\end{alignedat}
\]
Remark that we have equivalently rewritten \eqref{eq:BH:constraint},
\eqref{eq:BS:powerconstraint} and \eqref{eq:antenna:powerconsumption}
by introducing $s_{b}$ to the right hand side of these constraints
so as to improve the proposed algorithm's efficiency. Specifically,
if $s_{b}=0$ we can skip examining the constraints involving $s_{b}$.
Because ${\cal S}$ is upper bounded by the power and fronthaul constraints,
it satisfies the normal and finite properties required by a DBRB algorithm.
Let $D=[{\bf a};{\bf b}]\in\mathbb{R}_{+}^{BK+2B+K}$ be the box such
that ${\cal S}\subseteq D$, where the upper and lower vertices of
$D$ are defined as ${\bf a}\triangleq[\underline{{\bf x}},\underline{{\bf s}},\underline{{\bf r}},\underline{{\bf t}}]$
and ${\bf b}\triangleq[\overline{{\bf x}},\overline{{\bf s}},\overline{{\bf r}},\overline{{\bf t}}]$,
respectively. Vertices in ${\bf a}$ and ${\bf b}$ are calculated
as follows. It is obvious that $\underline{s}_{b}=0,\overline{s}_{b}=1,\underline{x}_{b,k}=0,\overline{x}_{b,k}=1$.
We can immediately see that $r_{k}\geq\underline{r}_{k}=r_{0}$ due
to \eqref{eq:achievable rate} and
\begin{align*}
r_{k}\leq\overline{r}_{k} & =\min\{\bar{C}_{b},\log(1+|{\bf h}_{k}{\bf w}_{k}|^{2}/\sigma_{k}^{2})\}\\
 & \leq\min\{\bar{C}_{b},\log(1+B\bar{P}_{b}\|{\bf h}_{k}\|_{2}^{2}/\sigma_{k}^{2})\}
\end{align*}
as $|{\bf h}_{k}{\bf w}_{k}|^{2}\leq\|{\bf h}_{k}\|_{2}^{2}\|{\bf w}_{k}\|_{2}^{2}$
by the Cauchy-Schwarz inequality, and $\|{\bf w}_{k}\|_{2}^{2}\leq B\bar{P}_{b}$.
We also have $t_{b}\geq\underline{t}_{b}=0$ and $t_{b}\leq\overline{t}_{b}=I\sqrt{P_{\text{a}}}$.

As mentioned above, we can solve \eqref{Prob:EE:epi} by branching
over $({\bf x},{\bf s},{\bf r},{\bf t})$. Recall that branching and
reduction for binary variables ${\bf x}$ and ${\bf s}$ follow Remarks
\ref{rem:Boolean:branch} and \ref{rem:Boolean:reduce}. In bounding
step, because the objective $\eta$ is determined via $({\bf x},{\bf s},{\bf r},{\bf t})$,
the upper and lower bounds of $\eta$ over a specific box $V=[\underline{{\bf x}},\underline{{\bf s}},\underline{{\bf r}},\underline{{\bf t}};\overline{{\bf x}},\overline{{\bf s}},\overline{{\bf r}},\overline{{\bf t}}]\subset{\cal R}_{n}$
can be simply calculated as $\eta_{\text{L}}(V)\triangleq\frac{\sum_{k\in{\cal K}}\underline{r}_{k}}{\hat{f}_{\text{P}}(\overline{{\bf x}},\overline{{\bf s}},\overline{{\bf r}},\overline{{\bf t}})}$
and $\eta_{\text{U}}(V)\triangleq\frac{\sum_{k\in{\cal K}}\overline{r}_{k}}{\hat{f}_{\text{P}}(\underline{{\bf x}},\underline{{\bf s}},\underline{{\bf r}},\underline{{\bf t}})}$.
Note that we need to verify whether box $V$ potentially contains
a feasible beamforming solution to \eqref{Prob:EE:epi} before bounding.
For the considered problem, we provide a better way of computing the
lower and upper bounds, and checking the feasibility of candidate
box $V$ during the bounding process. In what follows, we present
modifications (compared to the generic framework) made in Algorithm
\ref{Alg. BRB} to improve its efficiency.

\subsubsection*{Improved Branching\label{subsec:Dimensions-for-Branching}}

Normally each entry of $({\bf x},{\bf s},{\bf r},{\bf t})$ is branched
at each iteration, and thus the total number of iterations may increase
quickly with the problem size. For \eqref{Prob:EE:epi}, it turns
out that we can skip branching on ${\bf t}$ while still guaranteeing
the convergence. In particular, let us consider the following SOCP
\begin{subequations}\label{Prob:SOCP:checkfeasible}
\begin{align}
\underset{{\bf w},{\bf u}}{\textnormal{minimize}} & \quad\sum_{b\in{\cal B}}\sum_{i=1}^{I}u_{b,i}\\
\text{subject to}\  & \mathbf{h}_{k}\mathbf{w}_{k}\geq\sqrt{(e^{\underline{r}_{k}}-1)({\textstyle \sum_{j\neq k}^{K}}|\mathbf{h}_{k}\mathbf{w}_{j}|^{2}+\sigma^{2})}\label{eq:QOS:lowerbound}\\
 & ||\tilde{\mathbf{w}}_{b,i}||_{2}\leq u_{b,i},\ \underline{s}_{b}\underline{t}_{b}\leq\sum_{i=1}^{I}u_{b,i}\leq\overline{s}_{b}\overline{t}_{b},b\in{\cal B}\label{eq:beam:antenna:epi}\\
 & ||\tilde{\mathbf{w}}_{b,i}||_{2}^{2}\leq\overline{s}_{b}P_{\text{a}},\ \|{\bf w}_{b,k}\|_{2}^{2}\leq\overline{x}_{b,k}\bar{P}_{b},b\in{\cal B}\label{eq:powerconsumption:antenna:epi}\\
 & \sum_{k\in{\cal K}}\|{\bf w}_{b,k}\|_{2}^{2}\leq\overline{s}_{b}\bar{P}_{b}\ \forall b\in{\cal B}
\end{align}
\end{subequations} which can be viewed as minimizing the power consumption
subject to minimum users' rate requirement $\underline{{\bf r}}$.
Let us denote by ${\bf u}^{\ast}$ the optimal solution if \eqref{Prob:SOCP:checkfeasible}
is feasible and ${\bf t}^{\ast}\triangleq\{t_{b}^{\ast}\}_{b}$ with
$t_{b}^{\ast}=\sum_{i=1}^{I}u_{b,i}^{\ast}$. Obviously ${\bf t}^{\ast}$
is the minimum power required to achieve $\underline{{\bf r}}$, and
it holds $\underline{{\bf t}}\leq{\bf t}^{\ast}$. Also, $t_{b}^{\ast}$
is unique solution because the objective in \eqref{Prob:SOCP:checkfeasible}
is the epigraph of the function $\sum_{b\in{\cal B}}\sum_{i=1}^{I}||\tilde{\mathbf{w}}_{b,i}||_{2}$
\cite[Chapter 3]{dattorro2010convex}. At this point, we can replace
$\underline{{\bf t}}$ by ${\bf t}^{\ast}$ to obtain a tighter lower
bound on ${\bf t}$. Thus, it is sufficient to only branch $({\bf x},{\bf s},{\bf r})$
as the lower bound on ${\bf t}$ is always improved with $\underline{{\bf r}}$.
The property significantly accelerates the convergence of the proposed
algorithm.

\subsubsection*{Improved Branching Order\label{subsec:First-Branching-Variable}}

Essentially, in each iteration of a DBRB algorithm we can randomly
select a variable to perform branching. Exploiting the specifics of
the considered problem, we can potentially reduce the computational
complexity if we opt to branch ${\bf s}$ first due to its dependency
on other factors. Intuitively, the number of active RRHs provides
the degree-of-freedoms that can make the desired data rate $\underline{{\bf r}}$
achievable. Moreover, we can immediately obtain $x_{b,k}=0,\forall k\in{\cal K}$
whenever $s_{b}=0$, implying that the effective dimension in $V$
is reduced by $K$ times. Therefore by first keeping branching on
${\bf s}$ until $\underline{{\bf s}}=\overline{{\bf s}}$ , we can
quickly remove combinations of $\{s_{b}\}_{b}$ infeasible to \eqref{Prob:EE:epi}.
This is done by solving \eqref{Prob:SOCP:checkfeasible} with given
$\overline{{\bf s}}$ and target rate $r_{0}$ for all users. Moreover,
since the length of ${\bf s}$ is much smaller than that of ${\bf x}$
in most of wireless communications applications, branching on ${\bf s}$
may take a relatively small number of iterations.

\subsubsection*{Improved Memory Requirement\label{subsec:Checking-Feasibility}}

A DBRB algorithm basically stores a sequence of boxes until an optimal
solution is found, which requires some memory capacity. To reduce
this memory requirement we can eliminate boxes that contain no feasible
solution. Recall that the feasible set of \eqref{Prob:EE:epi} is
determined by the users' rate requirement, power and fronthaul constraints.
It is easily seen that the rate and power feasibility of box $V$
is equivalent to solving problem \eqref{Prob:SOCP:checkfeasible}.
For fronthaul constraints, we have the following feasibility condition,
i.e., if the inequality below does not hold
\begin{equation}
\sum_{k\in{\cal K}}\underline{r}_{k}\leq\sum_{b\in{\cal B}}\overline{s}_{b}\bar{C}_{b}\label{eq:bh:feasibility}
\end{equation}
then $V$ contains no feasible solution. In fact, \eqref{eq:bh:feasibility}
is due to $\sum_{b\in{\cal B}}\overline{s}_{b}\bar{C}_{b}\geq\sum_{b\in{\cal B}}\sum_{k\in{\cal K}}x_{b,k}r_{k}\geq\sum_{k\in{\cal K}}r_{k}\sum_{b\in{\cal B}}x_{b,k}\geq\sum_{k\in{\cal K}}\underline{r}_{k}$
where the last inequality follows \eqref{eq:min:connection}. In Algorithm
\ref{Alg. BRB}, we check \eqref{eq:bh:feasibility} prior to \eqref{Prob:SOCP:checkfeasible}
for saving computational efforts. We remark that the computational
complexity of checking the feasibility of $V$ is dominated by solving
\eqref{Prob:SOCP:checkfeasible}, and is independent of the dimension
of binary variables.

\subsubsection*{Improved Bounds\label{subsec:Bounding-Computation}}

Using monotonicity to compute bounds as mentioned above is inefficient
for our considered problem. We now present a way to obtain tighter
bounds which can improve the convergence rate of Algorithm \ref{Alg. BRB}
in practice. First recall that $\hat{f}_{\text{P}}(\underline{{\bf x}},\underline{{\bf s}},\underline{{\bf r}},\underline{{\bf t}})=\sum_{b\in{\cal B}}(\tilde{\epsilon}\underline{t}_{b}+\Delta P\underline{s}_{b}+p_{\text{SP}}\sum_{k\in{\cal K}}\underline{x}_{b,k}\underline{r}_{k})+P_{\text{const}}$
and observe that the terms involving binary variables are zero if
$\underline{s}_{b}=0$ and $\underline{x}_{b,k}=0$ for some $b,k$,
whereas $\Delta P$ and $p_{\text{SP}}$, i.e., the power for operating
RRHs and signal processing circuits are much larger than the power
consumption on the PAs. Let us consider the following bound
\begin{equation}
\begin{alignedat}{1} & \underline{\hat{f}}_{\text{P}}({\bf x},{\bf s},{\bf r},{\bf t}^{\ast})\triangleq\sum_{b\in{\cal B}}\tilde{\epsilon}t_{b}^{\ast}+\Delta P\max\{1,\sum_{b\in{\cal B}}\underline{s}_{b}\}\\
 & \quad+p_{\text{SP}}\max\{\sum_{k\in{\cal K}}\underline{r}_{k},\sum_{b\in{\cal B}}\sum_{k\in{\cal K}}\underline{x}_{b,k}\underline{r}_{k}\}+P_{\text{const}}
\end{alignedat}
\end{equation}
in which the first term is a result of solving \eqref{Prob:SOCP:checkfeasible}
(if feasible); the second term is due to the fact that at least one
RRH is active for transmission; the third term is achieved by $\sum_{b\in{\cal B}}(\sum_{k\in{\cal K}}x_{b,k}r_{k})\geq\sum_{k\in{\cal K}}r_{k}(\sum_{b\in{\cal B}}x_{b,k})\geq\sum_{k\in{\cal K}}r_{k}$.
Obviously, $\hat{f}_{\text{P}}(\underline{{\bf x}},\underline{{\bf s}},\underline{{\bf r}},\underline{{\bf t}})\leq\underline{\hat{f}}_{\text{P}}({\bf x},{\bf s},{\bf r},{\bf t}^{\ast})$
and replacing $\hat{f}_{\text{P}}(\underline{{\bf x}},\underline{{\bf s}},\underline{{\bf r}},\underline{{\bf t}})$
by $\underline{\hat{f}}_{\text{P}}({\bf x},{\bf s},{\bf r},{\bf t}^{\ast})$
does not remove any feasible solution. A tighter upper bound on $\eta$
over $V$ can be recalculated as
\begin{equation}
\eta_{\text{U}}(V)=\frac{\sum_{k\in{\cal K}}\overline{r}_{k}}{\underline{\hat{f}}_{\text{P}}({\bf x},{\bf s},{\bf r},{\bf t}^{\ast})}.\label{eq:eta:upbound}
\end{equation}
Similarly, suppose $(\hat{{\bf x}},\hat{{\bf s}},\underline{{\bf r}},\hat{{\bf t}})_{V}$
to be some feasible point within $V$. We can easily check that $\hat{f}_{\text{P}}(\hat{{\bf x}},\hat{{\bf s}},\underline{{\bf r}},\hat{{\bf t}})_{V}\leq\hat{f}_{\text{P}}(\overline{{\bf x}},\overline{{\bf s}},\overline{{\bf r}},\overline{{\bf t}})$
due to the monotonicity property of $\hat{f}_{\text{P}}({\bf x},{\bf s},{\bf r},{\bf t})$.
Then an improved lower bound on $\eta$ over $V$ can be obtained
as
\begin{equation}
\eta_{\text{L}}(V)=\frac{\sum_{k\in{\cal K}}\underline{r}_{k}}{\hat{f}_{\text{P}}(\hat{{\bf x}},\hat{{\bf s}},\underline{{\bf r}},\hat{{\bf t}})_{V}}.\label{eq:eta:lowbound}
\end{equation}
Remark that if $\eta_{\text{L}}(V)\geq\eta_{n}^{\text{best}}$ where
$\eta_{n}^{\text{best}}$ denotes the CBO at iteration $n$, we can
update $\eta_{\text{L}}(V)$ as the new CBO and then remove boxes
whose upper bounds are smaller than $\eta_{n}^{\text{best}}$ (see
Step \ref{remove} in Algorithm 1). Thus, obtaining a feasible point
is vital for improving the algorithm's efficiency. For this purpose
we present in the following a heuristic way.

\subsubsection*{Heuristic Method for Finding a Feasible Solution \label{subsec:Finding-a-feasible}}

We propose a simple trick which may quickly find a feasible solution
in $V$. It is worth noting that a feasible point $(\hat{{\bf x}},\hat{{\bf s}},\underline{{\bf r}},\hat{{\bf t}})_{V}$
of problem \eqref{Prob:EE:epi} must satisfy two conditions: $\underline{{\bf r}}$
is achievable by $(\hat{{\bf x}},\hat{{\bf s}},\hat{{\bf t}})_{V}$;
and
\begin{equation}
\hat{{\bf x}}\in\{{\bf x}\mid\sum_{b\in{\cal B}}x_{b,k}\geq1,k\in{\cal K},\sum_{k\in{\cal K}}x_{b,k}\underline{r}_{k}\leq\bar{C}_{b},b\in{\cal B}\}.\label{eq:feasible:cond}
\end{equation}
As can be easily seen, the feasible solution returned by solving \eqref{Prob:SOCP:checkfeasible}
always satisfies the former condition. Thus, our idea is to extract
$\hat{{\bf x}}$ from the optimal point of \eqref{Prob:SOCP:checkfeasible}
and verify \eqref{eq:feasible:cond}. Specifically, we can compute
$\hat{{\bf x}}$ by setting $\hat{x}_{b,k}=0$ if $\|{\bf w}_{b,k}^{\ast}\|_{2}=0$
and vice versa $\hat{x}_{b,k}=1$ if $\|{\bf w}_{b,k}^{\ast}\|_{2}>0$
where ${\bf w}^{\ast}$ is an optimal solution obtained by solving
\eqref{Prob:SOCP:checkfeasible}.

\subsection*{Convergence Analysis of Algorithm \ref{Alg. BRB}}

Algorithm 1 is guaranteed to yield a globally optimal solution of
\eqref{Prob:Gen:Problem} which can be justified following the same
arguments in the convergence analysis of generic DBRB \cite{tuy2006discrete}.
Specifically, we first recall that the branching and reduction operations
follow the same manner as in \cite{tuy2005monotonic,tuy2006discrete}.
These guarantee that the upper and lower bounds of $\eta$ in each
box are always improved after every iteration (branching rule), and
that no feasible point in a box being lost (reduction operations)\cite{tuy2006discrete}.
On the other hand, during the bounding step, it is easy to check that
the feasibility conditions (i.e.,\ \eqref{Prob:SOCP:checkfeasible}
and \eqref{eq:bh:feasibility}) and the calculation of tighter upper
and lower bounds (i.e.,\ \eqref{eq:eta:upbound} and \eqref{eq:eta:lowbound})
do not eliminate any feasible point, and the upper bound \eqref{eq:eta:upbound}
(resp. lower bound \eqref{eq:eta:lowbound}) is non-increasing (resp.
non-decreasing). We note that the feasible set is upper bounded by
the power and fronthaul constraints, and lower bounded by the users'
QoS constraints. Therefore, following the proof of \cite[Theorem 17]{tuy2006discrete},
Algorithm \ref{Alg. BRB} generates a sequence of boxes such that
the gap between the upper bound and lower bound is guaranteed to converge
to a single point, which is a globally optimal solution of \eqref{Prob:EE:epi}.
Recall that \eqref{Prob:Gen:Problem} and \eqref{Prob:EE:epi} is
optimally equivalent, thus Algorithm \ref{Alg. BRB} achieves globally
optimal solution of \eqref{Prob:Gen:Problem}.

\section{Suboptimal Designs}

In general a global optimization algorithm often takes enormous complexity
to output a solution. In this section, we propose two sub-optimal
approaches that are more practically appealing.

\subsection{Penalty Method}

In the first method, a binary variable is equivalently represented
by a set of continuous functions and then a penalty method is applied.
Note that we can rewrite \eqref{Prob:Gen:Problem} as\begin{subequations}\label{Prob:EE:epi:2}
\begin{align}
\underset{\substack{\substack{\eta,t,{\bf w},{\bf s},{\bf x},\\
{\bf r},{\bf g},{\bf q},\bm{\vartheta}
}
}
}{\text{maximize}} & \quad\eta\\
\text{subject to}\  & \eta t\leq\sum_{k\in{\cal K}}r_{k}\label{eq:EE:epi:2}\\
 & t\geq\tilde{f}_{\text{P}}({\bf w},{\bf x},{\bf s},\bm{\vartheta})\label{eq:power:epi}\\
 & \log(1+g_{k})\geq r_{k}\ \forall k\in{\cal K}\label{eq:rate:low:epi}\\
 & q_{k}\geq\|[\sigma_{k},\ \{\mathbf{h}_{k}\mathbf{w}_{j}\}_{j\in{\cal K}\backslash k}]\|_{2}^{2},\forall k\in{\cal K}\label{eq:interference:low:epi}\\
 & q_{k}g_{k}\leq|\mathbf{h}_{k}\mathbf{w}_{k}|^{2}\ \forall k\in{\cal K}\label{eq:SINR:low:epi}\\
 & \sum_{k\in{\cal K}}x_{b,k}r_{k}\leq\vartheta_{b},\ \vartheta_{b}\in[0,\bar{C}_{b}],\ \forall b\in{\cal B}\label{eq:BH:constraint:epi}\\
 & \eqref{eq:QoS:constraint},\eqref{eq:BS:powerconstraint}-\eqref{eq:Boolean:constraint}
\end{align}
\end{subequations} where $\eta$, $t$, ${\bf g}\triangleq\{g_{k}\}_{k}$,
${\bf q}\triangleq\{q_{k}\}_{k}$, $\bm{\vartheta}\triangleq\{\vartheta_{b}\}_{b}$
are newly introduced slack variables, and $\tilde{f}_{\text{P}}({\bf w},{\bf x},{\bf s},\bm{\vartheta})\triangleq\sum_{b\in{\cal B}}(\sum_{i=1}^{I}\tilde{\epsilon}||\tilde{\mathbf{w}}_{b,i}||_{2}+\Delta Ps_{b}+p_{\text{SP}}\vartheta_{b})+P_{\text{const}}$.
We can further reformulate \eqref{Prob:EE:epi:2} as\begin{subequations}\label{Prob:EE:epi:3}
\begin{align}
\underset{\substack{\substack{\eta,t,{\bf w},{\bf s},{\bf x},\\
{\bf r},{\bf g},{\bf q},\bm{\vartheta}
}
}
}{\text{maximize}} & \quad\eta\\
\text{subject to}\  & (\eta+t)^{2}\leq\|[\eta,\ t]\|_{2}^{2}+2\sum_{k\in{\cal K}}z_{k}\label{eq:EE:epi:3}\\
 & (q_{k}+g_{k})^{2}\leq\|q_{k},g_{k},\sqrt{2}\mathbf{h}_{k}\mathbf{w}_{k}]\|_{2}^{2},\ k\in{\cal K}\label{eq:SINR:low:epi-1}\\
 & \sum_{k\in{\cal K}}(x_{b,k}+r_{k})^{2}\leq\|[\{x_{b,k}\}_{k},\{r_{k}\}_{k}]\|_{2}^{2}+2\vartheta_{b}\label{eq:BH:constraint:epi-1}\\
 & \eqref{eq:QoS:constraint},\eqref{eq:BS:powerconstraint}-\eqref{eq:Boolean:constraint},\eqref{eq:power:epi}-\eqref{eq:interference:low:epi}.
\end{align}
\end{subequations}Clearly \eqref{Prob:EE:epi:3} maintains the feasible
set of \eqref{Prob:Gen:Problem}. To invoke continuous optimization,
we now represent binary variables ${\bf x}$ and ${\bf s}$ by a continuous
constraint. To this end, we can use the well-known relaxation of binary
variables which is given as \cite[Section 1]{tuy2006discrete}
\begin{equation}
x_{b,k}\in\{0,1\},\forall b,k\Leftrightarrow\sum_{b\in{\cal B},k\in{\cal K}}x_{b,k}^{2}-x_{b,k}\geq0,\ x_{b,k}\in[0,1].\label{eq:Boolean:exact}
\end{equation}
The above representation is justified by the fact that $x_{b,k}^{2}-x_{b,k}<0$
for $x_{b,k}\in(0,1)$. We note that $s_{b}$ is automatically binary
when $x_{b,k}$ is so, which is due to \eqref{eq:selection:relation}.
Thus we can simply relax $s_{b}\in[0,1]$ and equivalently rewrite
\eqref{Prob:EE:epi:3} as
\begin{equation}
\begin{alignedat}{1}\max_{\bm{\Omega}\in{\cal S}_{\text{c}}\cap{\cal S}_{\text{nc}}}\ \eta\quad\text{subject to}\quad\{ & \eqref{eq:Boolean:exact},\ s_{b}\in[0,1]\}\end{alignedat}
\label{Prob:EE:epi:4}
\end{equation}
where $\bm{\Omega}\triangleq\{\eta,t,{\bf w},{\bf s},{\bf x},{\bf r},{\bf g},{\bf q},\bm{\vartheta}\}$
and
\begin{align*}
{\cal S}_{\text{c}} & \triangleq\{\bm{\Omega}|\eqref{eq:QoS:constraint},\eqref{eq:BS:powerconstraint}-\eqref{eq:selection:relation},\eqref{eq:power:epi}-\eqref{eq:interference:low:epi}\}\\
{\cal S}_{\text{nc}} & \triangleq\{\bm{\Omega}|\eqref{eq:EE:epi:3}-\eqref{eq:BH:constraint:epi-1}\}
\end{align*}
which are the set of convex and nonconvex constraints of \eqref{Prob:EE:epi:4},
respectively. From this point onwards, $x_{b,k}$'s and $s_{b}$'s
are understood to be continuous over $[0,1]$. Now \eqref{Prob:EE:epi:4}
is a continuous nonconvex problem, for which one can basically apply
the SCA method to solve. However, finding an initial point of the
iterative process is usually difficult. To overcome the issue, we
apply a penalty method which results in the following regularized
problem
\begin{equation}
\begin{alignedat}{1}\max_{\substack{\bm{\Omega}\in{\cal S}_{\text{c}}\cap{\cal S}_{\text{nc}}}
} & \psi(\bm{\Omega},\alpha,\xi)\triangleq\eta+\alpha\sum_{b\in{\cal B},k\in{\cal K}}(x_{b,k}^{2}-x_{b,k})\\
 & \qquad\qquad\qquad+\xi\sum_{b\in{\cal B}}\min\{0,\bar{C}_{b}-\vartheta_{b}\}
\end{alignedat}
\label{Prob:EE:epi:5}
\end{equation}
where $\alpha,\xi>0$ are the penalty parameters. Intuitively, the
second term in $\psi(\bm{\Omega},\alpha,\xi)$ represents the cost
when $x_{b,k}$'s are not binary, while the last term represents the
cost when the fronthaul constraints are violated. Our expectation
is that solving \eqref{Prob:EE:epi:5} will eventually produce binary
solutions. In this regard we replace \eqref{eq:beam:powerconstraint}
by
\begin{equation}
\|{\bf w}_{b,k}\|_{2}^{2}\leq x_{b,k}^{q}\bar{P}_{b}.\label{eq:beam:power:replace}
\end{equation}
We can check that \eqref{eq:beam:power:replace} is equivalent to
\eqref{eq:beam:powerconstraint} for $x_{b,k}\in\{0,1\}$. To appreciate
the above maneuver, let $\mathcal{F}_{q}$ denote the feasible set
of \eqref{Prob:EE:epi:5} when \eqref{eq:beam:powerconstraint} is
replaced by \eqref{eq:beam:power:replace} and $\tilde{\eta}_{q}$
is the resulting optimal objective. For $x_{b,k}\in[0,1]$ it is clear
that $x_{b,k}^{q}\geq x_{b,k}^{q+1}$ for any $q>0$, meaning
\begin{equation}
\mathcal{F}_{q+1}\subseteq\mathcal{F}_{q}\subseteq\cdots\subseteq\mathcal{F}_{1}\triangleq{\cal S}_{\text{c}}\cap{\cal S}_{\text{nc}}
\end{equation}
and thus
\begin{equation}
\tilde{\eta}^{\ast}\leq\tilde{\eta}_{q+1}\leq\tilde{\eta}_{q}\leq\cdots\leq\tilde{\eta}_{1}
\end{equation}
where $\tilde{\eta}^{\ast}$ is the optimal value of \eqref{Prob:EE:epi:5}
for $x_{b,k}\in\{0,1\}$. The above inequality simply implies that
a tighter continuous relaxation can be obtained with higher values
of $q$. However we also note that \eqref{eq:beam:power:replace}
for $q>1$ is noncovex and thus it has not been used in the development
of the proposed global optimization algorithm.

Now we can apply the SCA to solve \eqref{Prob:EE:epi:5}. In the light
of the SCA principle \cite{MarksWright:78:AGenInnerApprox}, the nonconvex
constraints in ${\cal S}_{\text{nc}}$ and \eqref{eq:beam:power:replace}
can be approximated as
\begin{gather}
(\eta+t)^{2}\leq2[\eta^{n},t^{n}][\eta,t]^{T}-\|[\eta^{n},t^{n}]\|_{2}^{2}+2\sum_{k\in{\cal K}}r_{k}\label{eq:EE:epi:approx}\\
\begin{alignedat}{1}(q_{k}+g_{k})^{2} & \leq2\Re([q_{k}^{n},g_{k}^{n},\sqrt{2}\mathbf{h}_{k}\mathbf{w}_{k}^{n}][q_{k},g_{k},\sqrt{2}\mathbf{h}_{k}\mathbf{w}_{k}]^{H})\\
 & \qquad-\|[q_{k}^{n},g_{k}^{n},\sqrt{2}\mathbf{h}_{k}\mathbf{w}_{k}^{n}]\|_{2}^{2},\ \forall k
\end{alignedat}
\label{eq:SINR:low:epi:approx}\\
\begin{alignedat}{1}\sum_{k\in{\cal K}}(x_{b,k}+r_{k})^{2} & \leq2[\{x_{b,k}^{n}\}_{k},\{r_{k}^{n}\}_{k}][\{x_{b,k}\}_{k},\{r_{k}\}_{k}]^{T}\\
 & \ -\|[\{x_{b,k}\}_{k},\{r_{k}\}_{k}]\|_{2}^{2}+2\vartheta_{b},\ \forall b
\end{alignedat}
\label{eq:BH:constraint:epi:approx}\\
\begin{alignedat}{1}\|{\bf w}_{b,k}\|_{2}^{2} & \leq(q(x_{b,k}^{n})^{q-1}x_{b,k}+(1-q)(x_{b,k}^{n})^{q})\bar{P}_{b}\end{alignedat}
,\ \forall b,k.\label{eq:beam:power:replace:approx}
\end{gather}
Herein, the superscript $n$ denotes the iteration. Moreover, we also
convexify $\psi(\bm{\Omega},\alpha,\xi)$ using the first order as
$\psi(\bm{\Omega},\alpha,\xi;\bm{\Omega}^{n})\triangleq\eta+\alpha\sum_{b\in{\cal B},k\in{\cal K}}(2x_{b,k}x_{b,k}^{n}-(x_{b,k}^{n})^{2}-x_{b,k})+\xi\sum_{b\in{\cal B}}\min\{0,\bar{C}_{b}-\vartheta_{b}\}$.
\begin{algorithm}[tb]
\caption{Proposed method for solving \eqref{Prob:EE:epi:2}}
\label{Alg.SCA}

\begin{algorithmic}[1]

\STATE \textbf{Initialization:} Set $n:=0$,\textbf{ }choose initial
values for $\bm{\Omega}^{0}$ and set $\alpha^{0}$ small \\
\REPEAT[$n:=n+1$]

\STATE{Solve \eqref{Prob:EE:approx} and achieve $\bm{\Omega}^{\ast}$}

\STATE{Update $\bm{\Omega}^{n}:=\bm{\Omega}^{\ast}$ }

\STATE{Update $\alpha^{n}:=\min\{\alpha_{\max};\alpha^{n-1}+\varepsilon\}$
for small $\varepsilon$ }\label{Update penalty}

\UNTIL{Convergence}

\end{algorithmic}
\end{algorithm}
 In summary, at iteration $n+1$ of the proposed method, we solve
the following approximate convex program of \eqref{Prob:EE:epi:5}
\begin{align}
\max_{\substack{\bm{\Omega}\in{\cal S}_{\text{c}}\backslash\eqref{eq:beam:powerconstraint}}
}\quad & \psi(\bm{\Omega},\alpha,\xi;\bm{\Omega}^{n})\quad\text{subject to}\quad\{\eqref{eq:EE:epi:approx}-\eqref{eq:beam:power:replace:approx}\}.\label{Prob:EE:approx}
\end{align}
The convergence of Algorithm \ref{Alg.SCA} can be proved following
the arguments in \cite[Section 2]{le2014dc}. We also refer the interested
reader to \cite{le2012exact,le2015dc,MarksWright:78:AGenInnerApprox}
for other convergence results.

An important point in Algorithm \ref{Alg.SCA} is that the value of
penalty parameter $\alpha$ is increased at each iteration, i.e.,
step \ref{Update penalty}. We note that a high value of $\alpha$
will encourage $x_{b,k}$ to take on binary values. The idea is to
start Algorithm \ref{Alg.SCA} with a small value of $\alpha$ to
focus on maximizing the original objective, and then increase $\alpha$
in subsequent iterations to force $x_{b,k}$ to be binary.

\subsection{$\ell_{0}$-Approximation Method}

In the second suboptimal method, we view the problem of RRH selection
and RRH-user association as finding a sparse solution of beamformer
vector ${\bf w}$. In particular, no binary variables are introduced
to formulate the considered problem. Instead, RRH selection and RRH-user
association are concluded from the values of beamformers. To clarify
this point, let us consider the inequality $\|{\bf w}_{b,k}\|_{2}\leq v_{b,k}$.
Then it is clear that RRH $b$ is switched off if $\sum_{k\in{\cal K}}v_{b,k}=0$,
and switched on if $\sum_{k\in{\cal K}}v_{b,k}>0$. In other words,
whether RRH $b$ is active or not is the step function of $\sum_{k\in{\cal K}}v_{b,k}$.
The central idea of the second proposed method is to approximate the
step function by a continuous function to which continuous optimization
can be applied. In fact there are many functions proposed in the literature
for this purpose in different contexts (see \cite{le2015dc} for further
discussions on approximations). For the considered problem, we find
the following approximation function is very efficient
\begin{equation}
\varphi_{\beta}(y)\triangleq\min\{1,\beta y\}=\begin{cases}
1 & \text{if }y\geq\frac{1}{\beta}\\
\beta y & \text{if otherwise}
\end{cases}\label{eq:ell-cap}
\end{equation}
where $\beta$ is the approximation parameter. In fact the above approximation
function is a special case of (nonconcave) piecewise linear function
presented in \cite{le2015dc,zhang2010analysis}, which is modified
to be concave for the purpose of applying the SCA later on. We can
easily see that $\varphi_{\beta}(y)$ well approximates the step function
when $\beta$ is sufficiently large. Based on the above discussion,
we formulate the joint design problem as \begin{subequations}\label{Prob:Gen:Problem-1}
\begin{align}
\underset{{\bf w},{\bf r},{\bf v}}{\text{maximize}}\  & \quad\frac{\sum_{k\in{\cal K}}r_{k}}{\check{f}_{\text{P}}({\bf w},{\bf r},{\bf v})}\\
\text{subject to}\  & \|{\bf w}_{b,k}\|_{2}\leq v_{b,k},\ \sum_{k\in{\cal K}}v_{b,k}^{2}\leq\bar{P}_{b},\ \forall b\in{\cal B}\label{eq:power:slackvar}\\
 & \sum_{k\in{\cal K}}\varphi_{\beta}(v_{b,k})r_{k}\leq\bar{C}_{b},\ \forall b\in{\cal B}\label{eq:BH:constraint:sparsity}\\
 & \eqref{eq:achievable rate},\eqref{eq:QoS:constraint},\eqref{eq:per-antenna:powerconstraint}
\end{align}
\end{subequations} where ${\bf v}\triangleq\{v_{b,k}\}$ and $\check{f}_{\text{P}}({\bf w},{\bf r},{\bf v})\triangleq\sum_{b\in{\cal B}}\bigl(\sum_{i=1}^{I}\tilde{\epsilon}||\tilde{\mathbf{w}}_{b,i}||_{2}+\Delta P\varphi_{\beta}(\sum_{k\in{\cal K}}v_{b,k})+p_{\text{SP}}\sum_{k\in{\cal K}}\varphi_{\beta}(v_{b,k})r_{k}\bigr)+P_{\text{const}}$.
We note that \eqref{Prob:Gen:Problem-1} is still nonconvex but it
has fewer optimization variables than \eqref{Prob:Gen:Problem}. Next
we rewrite \eqref{Prob:Gen:Problem-1} as \begin{subequations}\label{Problem:EE:sparsity}
\begin{align}
\underset{\substack{\eta,t,{\bf w},{\bf r},{\bf v}\\
{\bf g},{\bf q},\bm{\vartheta},\bm{\mu},\bm{\nu}
}
}{\text{maximize}} & \qquad\eta\\
\text{subject to}\  & \begin{alignedat}{1}t\geq & \sum_{b\in{\cal B}}(\sum_{i=1}^{I}\tilde{\epsilon}||\tilde{\mathbf{w}}_{b,i}||_{2}+\Delta P\nu_{b}+p_{\text{SP}}\vartheta_{b})+P_{\text{const}}\end{alignedat}
\label{eq:power:epi:sparsity}\\
 & \sum_{k\in{\cal K}}\mu_{b,k}r_{k}\leq\vartheta_{b},\ \vartheta_{b}\in[0,\bar{C}_{b}],\ \forall b\in{\cal B}\label{eq:BH:usage:epi:3}\\
 & \mu_{b,k}\geq\varphi_{\beta}(v_{b,k}),\ \forall b\in{\cal B},k\in{\cal K}\label{eq:BS-user:epi}\\
 & \nu_{b}\geq\varphi_{\beta}({\textstyle \sum_{k\in{\cal K}}}v_{b,k}),\ \forall b\in{\cal B}\label{eq:BS:epi}\\
 & \eqref{eq:QoS:constraint},\eqref{eq:per-antenna:powerconstraint},\eqref{eq:rate:low:epi},\eqref{eq:interference:low:epi},\eqref{eq:EE:epi:3},\eqref{eq:SINR:low:epi-1},\eqref{eq:power:slackvar}
\end{align}
\end{subequations} where $\bm{\mu}\triangleq\{\mu_{b,k}\}_{b,k}$
and $\bm{\nu}\triangleq\{\nu_{b}\}_{b}$, and the introduction of
$\eta,t,{\bf g},{\bf q},\bm{\vartheta}$ follows exactly the same
arguments as those in the previous subsection. For the ease of description,
we define
\begin{align*}
\tilde{{\cal S}}_{\text{c}} & \triangleq\{\tilde{\bm{\Omega}}|\eqref{eq:QoS:constraint},\eqref{eq:per-antenna:powerconstraint},\eqref{eq:rate:low:epi},\eqref{eq:interference:low:epi},\eqref{eq:power:slackvar},\eqref{eq:power:epi:sparsity}\}\\
\tilde{{\cal S}}_{\text{nc}} & \triangleq\{\tilde{\bm{\Omega}}|\eqref{eq:EE:epi:3},\eqref{eq:SINR:low:epi-1},\eqref{eq:BH:usage:epi:3}-\eqref{eq:BS:epi}\}
\end{align*}
where $\tilde{\bm{\Omega}}\triangleq\{\eta,t,{\bf w},{\bf r},{\bf v},{\bf g},{\bf q},\bm{\vartheta},\bm{\mu},\bm{\nu}\}$.
Note that $\tilde{{\cal S}}_{\text{c}}$ and $\tilde{{\cal S}}_{\text{nc}}$
are the convex and nonconvex parts of \eqref{Problem:EE:sparsity},
respectively. Now the application of SCA to solve \eqref{Problem:EE:sparsity}
is straightforward. The nonconvex constraints \eqref{eq:EE:epi:3},
\eqref{eq:SINR:low:epi-1} and \eqref{eq:BH:usage:epi:3} in $\tilde{{\cal S}}_{\text{nc}}$
can be convexified using the same way as done in the previous subsection,
given in \eqref{eq:EE:epi:approx}\textendash \eqref{eq:BH:constraint:epi:approx}.
Convex approximation of $\varphi_{\beta}(y)$ deserves a remark. Note
that $\varphi_{\beta}(y)$ is concave and continuous but not smooth
at $y=\frac{1}{\beta}$. However we can use the sub-differential of
$\varphi_{\beta}(y)$ to derive a convex upper bound. It is easy to
check that a subgradient of $\varphi_{\beta}(y)$ is given by
\begin{equation}
\partial\varphi_{\beta}(y)=\begin{cases}
0 & \text{if }y\geq\frac{1}{\beta}\\
\beta & \text{if otherwise}
\end{cases}
\end{equation}
and thus we can approximate \eqref{eq:BS-user:epi} and \eqref{eq:BS:epi}
as\begin{subequations}
\begin{align}
\mu_{b,k} & \geq\bar{\varphi}_{\beta}(v_{b,k};v_{b,k}^{n})\label{eq:BS-user:epi:approx}\\
\nu_{b} & \geq\bar{\varphi}_{\beta}({\textstyle \sum_{k\in{\cal K}}}v_{b,k};{\textstyle \sum_{k\in{\cal K}}}v_{b,k}^{n})\label{eq:BS:epi:approx}
\end{align}
\end{subequations}where $\bar{\varphi}_{\beta}(y;y^{n})\triangleq\begin{cases}
1 & \text{if }y^{n}\geq\frac{1}{\beta}\\
\beta y & \text{otherwise}
\end{cases}$. Finally, we arrive at the approximate convex program of problem
\eqref{Problem:EE:sparsity}, i.e.,
\begin{algorithm}[tb]
\caption{Proposed method for solving \eqref{Prob:Gen:Problem-1}}
\label{Alg.SCA-1}

\begin{algorithmic}[1]

\STATE \textbf{Initialization:} Set $n:=0$,\textbf{ }choose initial
values for $\tilde{\bm{\Omega}}^{0}$ and set $\beta^{0}$ small \\
\REPEAT[$n:=n+1$]

\STATE{Solve \eqref{Prob:EE:sparsity:approx} and achieve $\tilde{\bm{\Omega}}^{\ast}$}

\STATE{Update $\tilde{\bm{\Omega}}^{n}:=\tilde{\bm{\Omega}}^{\ast}$
}

\STATE{Update $\beta^{n}:=\min\{\beta_{\max};\beta^{n-1}+\varepsilon\}$
for small $\varepsilon$ }\label{Update penalty-1}

\UNTIL{Convergence and output $\tilde{\bm{\Omega}}^{\ast}$}

\end{algorithmic}
\end{algorithm}
\begin{align}
\max_{\substack{\tilde{\bm{\Omega}}\in\tilde{{\cal S}}_{\text{c}}}
}\quad & \eta\quad\text{subject to}\ \{\eqref{eq:EE:epi:approx}-\eqref{eq:BH:constraint:epi:approx},\eqref{eq:BS-user:epi:approx},\eqref{eq:BS:epi:approx}\}.\label{Prob:EE:sparsity:approx}
\end{align}
We describe the second proposed suboptimal method in Algorithm \ref{Alg.SCA-1}\textcolor{blue}{.
}Similar to Algorithm \ref{Alg.SCA}, the approximation parameter
$\beta$ is also updated after each iteration. The idea is the same
as $\beta$ is viewed to provide the tightness of the binary approximation
function \eqref{eq:ell-cap}. In Algorithm \algnew~we start with
a small value of $\beta$ and then increase $\beta$ after each iteration.
Numerical results provided in the next section demonstrate the impact
of updating $\beta$. To avoid the problem of the initial guess, we
can add the penalty of violating the fronthaul constraints to the
objective of \eqref{Prob:EE:sparsity:approx} similarly as with Algorithm
\ref{Alg.SCA}. Convergence of Algorithm \ref{Alg.SCA-1} is guaranteed,
which is discussed in Appendix B. It is worth mentioning that the
achieved limit point is not ensured to hold the first-order optimality
of \eqref{Problem:EE:sparsity} since the approximation of the step
function is not smooth.

\subsection{Second-order-cone Representation}

This subsection presents a more efficient way to treat \eqref{eq:rate:low:epi}.
First we remark that \eqref{eq:rate:low:epi} is indeed a convex constraint
and thus convex approximation is not required. However, since \eqref{eq:rate:low:epi}
involves an exponential cone, \eqref{Prob:EE:approx} and \eqref{Prob:EE:sparsity:approx}
are generic nonlinear programs, while other constraints are SOC presentable.
This prevents us from exploiting powerful conic convex solvers such
as MOSEK or GUROBI. To take full advantage of these solvers we will
also approximate \eqref{eq:rate:low:epi} using the SCA framework.
More explicitly, we will approximate $\log(1+g_{k})$ by a lower bound
that makes the resulting constraint SOC representable. To this end
we recall the following inequality
\begin{equation}
\log(1+g_{k})\geq g_{k}(1+g_{k})^{-1}.\label{eq:log:ineq}
\end{equation}
Substituting $g_{k}$ in both sides of \eqref{eq:log:ineq} by $\frac{g_{k}-g_{k}^{n}}{g_{k}^{n}+1}$
results in
\begin{equation}
\log(1+g_{k})\geq\log(1+g_{k}^{n})+(g_{k}-g_{k}^{n})(1+g_{k})^{-1}.\label{eq:log:approx}
\end{equation}
Now we can approximate \eqref{eq:rate:low:epi} as
\begin{equation}
\log(1+g_{k}^{n})+(g_{k}-g_{k}^{n})(1+g_{k})^{-1}\geq r_{k}.\label{eq:rate:approx}
\end{equation}
The above constraint can be reformulated as an SOC constraint as
\begin{multline}
\|2\sqrt{1+g_{k}^{n}},\ \log(1+g_{k}^{n})-r_{k}-g_{k}\|_{2}\\
\leq\ \log(1+g_{k}^{n})-r_{k}+g_{k}+2,\ \forall k\in{\cal K}.\label{eq:log-approx-SOCP}
\end{multline}
Using \eqref{eq:log-approx-SOCP}, the convex program obtained at
each iteration of Algorithms \ref{Alg.SCA} and \algnew~is an SOCP
which is much easier to solve.

\subsection{Complexity Analysis of Algorithms \ref{Alg.SCA} and \algnew}

We now discuss the worst-case per-iteration complexity of the two
proposed suboptimal algorithms. For Algorithm \ref{Alg.SCA}, the
SOCP consists of $2BKI+BK+2B+3K+2$ real-valued variables and $B(K+I)+2B+3K+2$
conic constraints. Thus, the per-iteration complexity for solving
the SOCP problems corresponding to Algorithms \ref{Alg.SCA} by path-following
interior-point method are $\mathcal{O}(\sqrt{B(K+I)}B^{3}K^{3}I^{3})$
\cite{BenNemi:book:LectModConv}. Similarly, the per-iteration complexity
for solving the SOCP in Algorithm \algnew, which contains $2BKI+2BK+2B+3K+2$
real-valued variables and $B(K+I)+2B+3K+1$ conic constraints, is
$\mathcal{O}(\sqrt{B(K+I)}B^{3}K^{3}I^{3})$ \cite{BenNemi:book:LectModConv}.

\section{Numerical Results\label{sec:Numerical-Results}}

In this section we provide numerical demonstration to evaluate the
effectiveness of the proposed methods. The simulation parameters shown
in Table \ref{Tab. 1} are used, unless mentioned otherwise. The channel
${\bf h}_{b,k}$ between RRH $b$ and user $k$ is assumed to be flat
fading which is generated following Gaussian distribution, i.e., ${\bf h}_{b,k}\sim\mathcal{CN}(0,\rho_{b,k}{\bf I}_{I}$),
where $\rho_{b,k}$ represents the large-scale fading and is calculated
as $\rho_{b,k}[\text{dB}]=30\log_{10}(d_{b,k})\text{ }+38+\mathcal{N}(0,8)$
($d_{b,k}$ is the distance in meters). $P_{\text{a}}$ is set to
be same for all antenna chains, and we take $\bar{P}=\bar{P}_{b}=IP_{\text{a}},\forall b$.
For the maximum fronthaul capacity, we set $\bar{C}_{b}=\bar{C},\forall b$.

We generate initial point $\bm{\Omega}^{0}$ for starting Algorithm
\ref{Alg.SCA} by solving the power minimization problem \eqref{Prob:SOCP:checkfeasible}
with selection vectors being fixed as $x_{b,k}^{0}=1$ and $s_{b}^{0}=1$
$\forall b,k$ to obtain ${\bf w}^{0}$; then the values for the remaining
variables are determined based on \eqref{eq:EE:epi:2}\textendash \eqref{eq:BH:constraint:epi}.
The initial point $\tilde{\bm{\Omega}}^{0}$ for starting Algorithm
\algnew\ is generated similarly with $\mu_{b,k}^{0}=1$ and $\nu_{b}^{0}=1$
$\forall b,k$. For the penalty parameters, we take $\xi=1$ and initialize
$\alpha^{0}=10^{-5}$ and $\beta^{0}=0.1$. Algorithms \ref{Alg.SCA}
and \algnew~are terminated when the increase in the objective between
two consecutive iterations is less than $10^{-6}$.
\begin{table}[tb]
\caption{Simulation Parameters}

\centering{}%
\begin{tabular}{c|c}
\hline
\noun{Parameters} & \noun{Value}\tabularnewline
\hline
Inter-RRH distance & $200$ m\tabularnewline
Active power for RRH and NU $P^{\text{active}}$ \cite{HowmuchEnergy-Auer,Dhaini:2014:EE-TDMA} & 10.65 W\tabularnewline
Sleep power for RRH and NU $P^{\text{sleep}}$\cite{HowmuchEnergy-Auer,Dhaini:2014:EE-TDMA} & 5.05 W\tabularnewline
Circuit power for user $P_{\text{ms}}$ & 0.1 W\tabularnewline
Max. power efficiency $\epsilon_{\max}$\cite{AmplifierMIMO-Persson} & 0.55\tabularnewline
Number of Tx antennas $N$ & 2\tabularnewline
Min. rate requirement $r_{0}$ & 1 nat/s/Hz\tabularnewline
Bandwidth & 10 MHz\tabularnewline
Noise power & -143 dBW\tabularnewline
\hline
\end{tabular}\label{Tab. 1}
\end{table}

\subsection{Convergence Results}

The first set of experiments examines the convergence behavior of
the proposed methods. We consider the network setting where $B=3$,
$K=4$, $\bar{P}=30$ dBm, $\bar{C}=10$ nats/s/Hz and $p_{\text{SP}}=10$
W/(Gnats/Hz).

\begin{figure}
\begin{centering}
\subfigure[Convergence of the upper and lower bounds.]{\label{Fig.3a}\enskip{}\includegraphics[width=0.8\columnwidth]{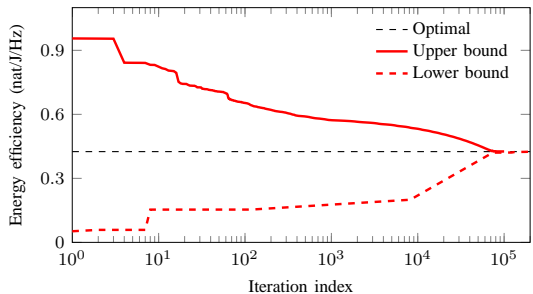}}\smallskip{}
\subfigure[Convergence speed with and without the proposed mofications.]{\label{Fig.3b}\includegraphics[width=0.84\columnwidth]{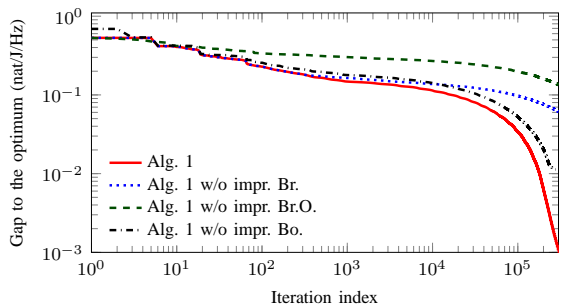}}
\par\end{centering}
\caption{Convergence behavior of Algorithm \ref{Alg. BRB} for one channel
realization with $B=3$, $K=4$, $\bar{P}=30$ dBm, $\bar{C}=10$
nats/s/Hz and $p_{\text{SP}}=10$ W/(Gnats/Hz).}
\label{Fig.2}
\end{figure}
The convergence performance of Algorithm \ref{Alg. BRB} for a random
channel realization is demonstrated in Fig. \ref{Fig.2}. Particularly,
Fig. \ref{Fig.3a} depicts the upper and lower bounds returned by
the algorithm. We can see that the bounds monotonically converge to
the optimal value. Fig. \ref{Fig.3b} shows the convergence speed
of Algorithm \ref{Alg. BRB} by the gap between the upper bound and
the optimal value over iterations. In this figure, we also provide
the performance of other schemes to confirm the effectiveness of the
proposed modifications made to the DBRB. Specifically, the schemes
labelled `\emph{w/o. impr. Br}.', `\emph{w/o. impr. Br.O}.', and
`\emph{w/o. impr. Bo.}' represent for Algorithm 1 without applying
improved branching, improved branching order and improved bounding,
respectively. The results clearly demonstrate that applying the proposed
modifications significantly improves the convergence performance.

\begin{figure}
\begin{centering}
\subfigure[Convergence behavior of the proposed suboptimal algorithms.]{\label{Fig.4a}\includegraphics[width=0.78\columnwidth]{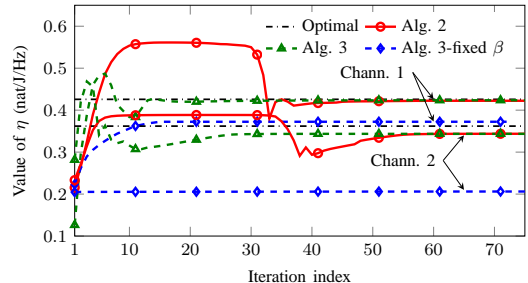}}\medskip{}
\subfigure[Gap to the binary of relaxed variables obtained by the proposed suboptimal algorithms.]{\label{Fig.4b}\includegraphics[width=0.8\columnwidth]{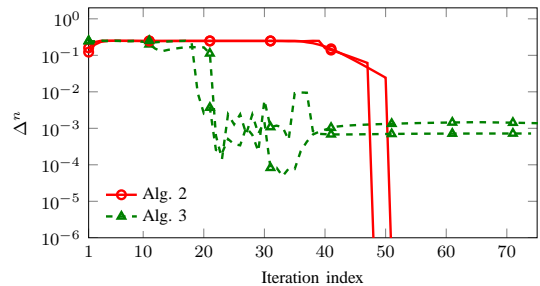}}
\par\end{centering}
\caption{Convergence behavior of the proposed suboptimal algorithms for two
random channel realizations with $B=3$, $K=4$, $\bar{P}=30$ dBm,
$\bar{C}=10$ nats/s/Hz and $p_{\text{SP}}=10$ W/(Gnats/Hz).}
\label{Fig.4}
\end{figure}
In Fig. \ref{Fig.4}, we show the convergence behavior of Algorithms
\ref{Alg.SCA} and \algnew\ for two random channel realizations.
In order to illustrate the advantages of updating parameter $\beta$
in Algorithm \algnew, we also provide the convergence results of
Algorithm \algnew~without updating $\beta$ dubbed as `\emph{Alg.
3-fixed $\beta$}'. For this scheme, we fix $\beta=1000$. Fig. \ref{Fig.4a}
shows the variation of $\eta$ over iterations. It is observed that
Algorithms \ref{Alg.SCA} and \algnew~converge to the points close
to the optimal values within a few tens of iterations. This behavior
proves that the proposed algorithms are fast convergent and effective
methods. Another observation is that, with a fixed $\beta$, Algorithm
\algnew\ converges very fast but results in poor performance. Whereas,
by updating $\beta$, the algorithm needs a bit more iterations to
achieve near-optimal performance. In Fig. \ref{Fig.4b}, we study
how close the obtained values of the relaxed variables are to 0 or
1. Let us define
\[
\Delta^{n}\triangleq\begin{cases}
\max_{b,k}\{x_{b,k}^{n}-(x_{b,k}^{n})^{2}\} & \textrm{for\ Algorithm\ 2}\\
\max_{b,k}\{\mu_{b,k}^{n}-(\mu_{b,k}^{n})^{2}\} & \textrm{for\ Algorithm\ 3}
\end{cases}
\]
and note that a smaller $\Delta^{n}$ indicates a closer gap between
$\{x_{b,k}^{n}\}_{b,k}$ (or $\{\mu_{b,k}^{n}\}_{b,k}$) and binary
values. As can be seen, $\Delta^{n}\approx0$ at convergence for Algorithm
\ref{Alg.SCA} which implies that the penalty method can achieve binary
solutions. On the other hand, although the $\ell_{0}$-approximation
method (i.e. Algorithm \algnew) cannot derive exact binary solutions
(for all relaxed variables), it still returns $\{\mu_{b,k}^{n}\}_{b,k}$
very close to 0 or 1 at convergence (the maximum gap is about $10^{-3}$).

For the solving time, the corresponding average per-iteration runtime
of solver MOSEK \cite{Mosek} with Algorithms 1, 2 and 3 are 0.006
s, 0.008 s, and 0.007 s, respectively. As can be seen, the per-iteration
runtime is relatively small due to the fact that only an SOCP is solved
in each iteration for all algorithms.

\subsection{EE  Comparison between Optimal and Suboptimal Algorithms}

\begin{figure}
\begin{centering}
\subfigure[Average EE versus $\bar{C}.$]{\label{Fig.5a}\includegraphics[width=0.8\columnwidth]{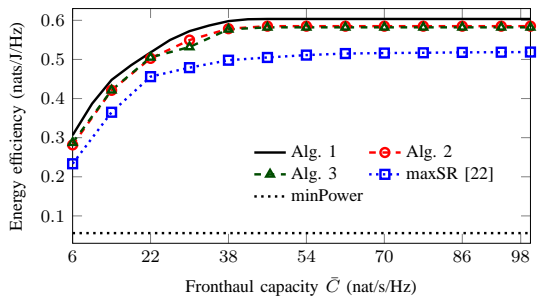}}\medskip{}
\subfigure[CDF of~the ratio of achieved EE (of the considered schemes)
to the optimal solution.]{\label{Fig.5b}\includegraphics[width=0.78\columnwidth]{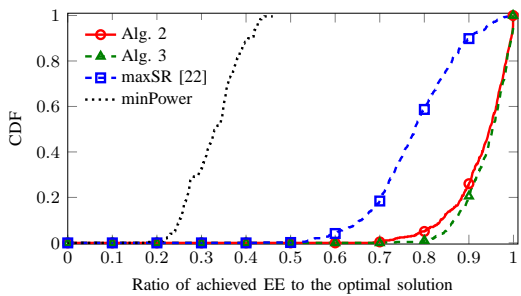}}
\par\end{centering}
\caption{Average performances of the considered schemes with $B=3$, $K=4$,
$\bar{P}=30$ dBm and $p_{\text{SP}}=10$ W/(Gnats/Hz).}
\label{Fig.5}
\end{figure}
In Fig. \ref{Fig.5}, we illustrate the effectiveness of the proposed
suboptimal algorithms by comparing their average EE performances with
that of Algorithm \ref{Alg. BRB} and the existing schemes, those
are sum rate maximization (maxSR) \cite{BinBinDai:15:Access:BeamformingBH}
and power consumption minimization (minPower) \cite{Dai2016:EECRAN,Hajisami:2017:ElasticNet,GroupGreenBeam-YShi}.
Fig. \ref{Fig.5a} plots the average EE of the considered schemes
as a function of the fronthaul capacity $\bar{C}$. It is seen that
the proposed methods remarkably outperform the existing schemes. The
important observation is that the gaps between the curves of the optimal
and suboptimal algorithms are really small in all cases of $\bar{C}$
demonstrating the validity of the proposed suboptimal schemes in terms
of the average EE. We can also observe that the performance of Algorithms
\ref{Alg.SCA} and \algnew~almost agree with each other. Fig. \ref{Fig.5b}
shows the cumulative distribution function (CDF) of the ratio of achieved
EE (of Algorithms \ref{Alg.SCA} and \algnew, maxSR, and minPower)
to the optimal solution. As can be observed, the probability that
Algorithms \ref{Alg.SCA} and \algnew~achieve more than 90\% of
the optimal values is up to 75\%. In the worst case, these schemes
also achieve about 70\% of the optimal performance. We can also see
that most of the solutions obtained by maxSR and minPower are far
from the optimal values.

\subsection{Performances of the Proposed Suboptimal Algorithms in Large Network
Settings}

In the following set of experiments, we consider a larger network
setting and evaluate the impacts of the fronthaul capacity, the signal
processing power and the dynamics of the PA's efficiency on the EE
performance. In particular, we evaluate the suboptimal methods in
a 7-cell wrap-around topology with $B=7$ RRHs in which a total of
$K=14$ users are randomly placed across the network's coverage.

\subsubsection{Impact of Fronthaul Capacity}

\begin{figure}[tp]
\begin{centering}
\includegraphics[width=0.8\columnwidth]{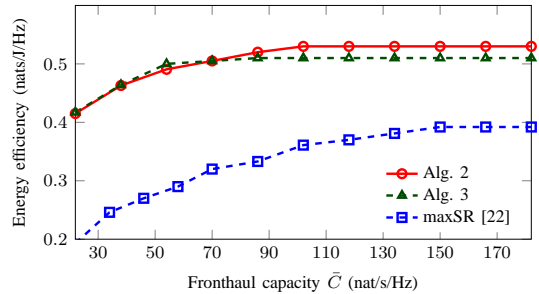}
\par\end{centering}
\caption{Average EE performance of the considered schemes versus $\bar{C}$
with $B=7$, $K=14$, $\bar{P}=30$ dBm and $p_{\text{SP}}=10$ W/(Gnats/Hz).}
\label{Fig.6}
\end{figure}
\begin{table*}
\centering{}\caption{Average number of served users per RRH and average number of serving
RRHs per user corresponding to the simulation results shown in Fig.
\ref{Fig.6}.}
\begin{tabular}{c|c||c|c|c|c|c|c|c|c|c|c|c}
\multicolumn{2}{c||}{$\bar{C}$ (nats/s/Hz)} & 22 & 30 & 38 & 46 & 54 & 66 & 70 & 78 & 86 & 118 & 150\tabularnewline
\hline
\hline
\multirow{2}{*}{Algorithm \ref{Alg.SCA}} & Num. of served users per RRH & 8.4 & 9.8 & 10.3 & 11.1 & 11.3 & 11.4 & 11.5 & 11.5 & 11.6 & 11.8 & 12.0\tabularnewline
\cline{2-13}
 & Num. of serving RRHs per user & 4.2 & 4.9 & 5.1 & 5.5 & 5.6 & 5.7 & 5.7 & 5.7 & 5.8 & 5.9 & 6.0\tabularnewline
\hline
\multirow{2}{*}{Algorithm \algnew} & Num. of served users per RRH & 8.0 & 8.6 & 8.9 & 9.4 & 9.6 & 9.7 & 9.8 & 9.8 & 9.8 & 9.8 & 9.8\tabularnewline
\cline{2-13}
 & Num. of serving RRHs per user & 4.0 & 4.3 & 4.5 & 4.7 & 4.8 & 4.9 & 4.9 & 4.9 & 4.9 & 4.9 & 4.9\tabularnewline
\hline
\multirow{2}{*}{maxSR \cite{BinBinDai:15:Access:BeamformingBH}} & Num. of served users per RRH & 11.3 & 12.0 & 12.3 & 12.6 & 13.0 & 13.6 & 13.8 & 13.8 & 13.8 & 14 & 14\tabularnewline
\cline{2-13}
 & Num. of serving RRHs per user & 5.6 & 6.0 & 6.1 & 6.3 & 6.5 & 6.8 & 6.9 & 6.9 & 6.9 & 7 & 7\tabularnewline
\hline
\end{tabular}\label{tab.3}
\end{table*}
Fig. \ref{Fig.6} shows the achieved EE of Algorithms \ref{Alg.SCA},
\algnew, and maxSR versus the fronthaul capacity $\bar{C}$. The
corresponding average number of served users per RRH and average number
of serving RRHs per user are provided in Table \ref{tab.3}. We can
see from Fig. \ref{Fig.6} that EE increases as $\bar{C}$ increases
for all considered schemes. However, after a certain large value of
$\bar{C}$, further increasing $\bar{C}$ does not change the performance.
This observation is consistent with that in Fig. \ref{Fig.5a}. The
result can be explained as follows. For the EE schemes, to increase
$\bar{C}$ is to expand feasible set of \eqref{Prob:Gen:Problem}.
When $\bar{C}$ is small, it is the primary constraint on the network
performances. Thus the expanded feasible set results in performance
improvement. When $\bar{C}$ is large enough, other constraints (e.g.
transmit power constraints) become the primary restriction on the
network performance. In this case, increasing $\bar{C}$ has no impact
on the objective value. For a physical interpretation, increasing
the fronthaul capacity allows a RRH to serve more users, i.e. the
number of RRHs cooperating to transmit data to a user increases (as
can be seen from Table \ref{tab.3}). This increases the cooperation
gain, and thus improves the system performance. When the fronthaul
capacity is large enough such that either the additional cooperation
gain provides no gain in the achieved performance or the full connection
(each user is served by all RRHs) is arrived, increasing the fronthaul
capacity does not change the performance. Therefore, in the large
fronthaul capacity regime, we can observe from the table that maxSR
arrives at full connection, since this topology provides the maximum
capacity for wireless transmission. On the other hand, for the EE
schemes, the average number of serving RRHs per user is smaller than
$B$ even when $\bar{C}$ is sufficiently large. This is because adding
more serving RRHs for a user degrades the EE performance, if the benefit
from the cooperation gain cannot compensate for the additional signal
processing power.

\subsubsection{Impact of Rate-dependent Signal Processing Power}

\begin{figure}
\begin{centering}
\includegraphics[width=0.79\columnwidth]{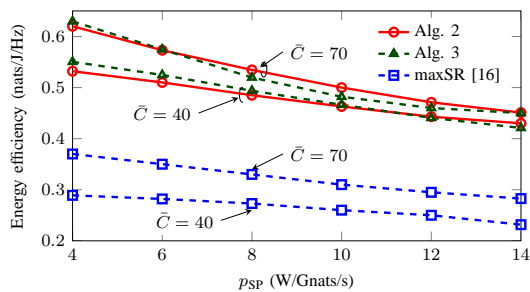}
\par\end{centering}
\caption{Average EE performance of the considered schemes versus $p_{\text{SP}}$
with $B=7$, $K=14$ and $\bar{P}=30$ dBm.}
\label{Fig.7}
\end{figure}

Fig. \ref{Fig.7} depicts the EE performance of the considered schemes
versus different values of $p_{\text{SP}}$. We recall that, for a
fixed data rate, a larger $p_{\text{SP}}$ leads to larger power consumed
in signal processing. As expected, the EE decreases when $p_{\text{SP}}$
increases for all considered schemes. For SRmax, the sum rate performance
is independent of $p_{\text{SP}}$. Thus, its EE performance is a
decreasing function of $p_{\text{SP}}$ due to the increase in the
total consumed power with respect to $p_{\text{SP}}$. The results
clearly show that parameter $p_{\text{SP}}$ has a significant impact
on the EE performance, indicating that the model of rate-dependent
signal processing power should be considered for proper EE designs
and evaluation.

\subsubsection{Impact of the Dynamics of PA's Efficiency}

\begin{figure}
\begin{centering}
\subfigure[Average EE]{\label{Fig.9a}\includegraphics[width=0.79\columnwidth]{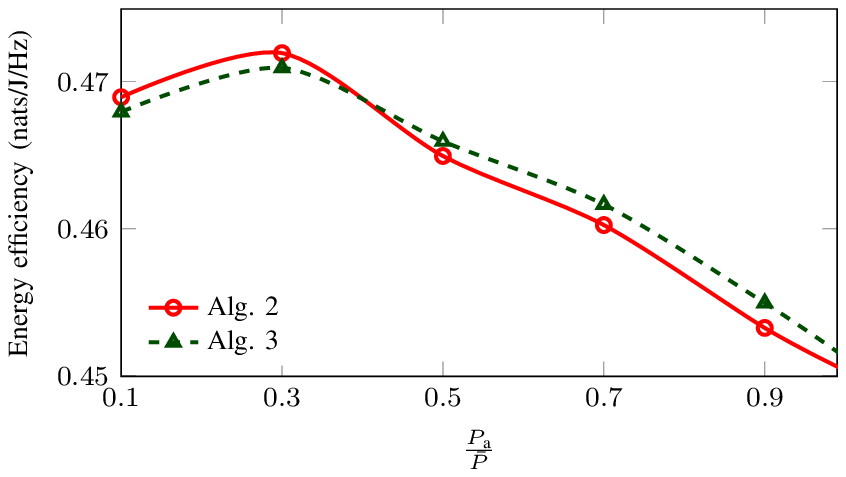}}\medskip{}
\subfigure[Average sum rate and total power consumption]{\label{Fig.9b}\hspace{3em}\includegraphics[width=0.86\columnwidth]{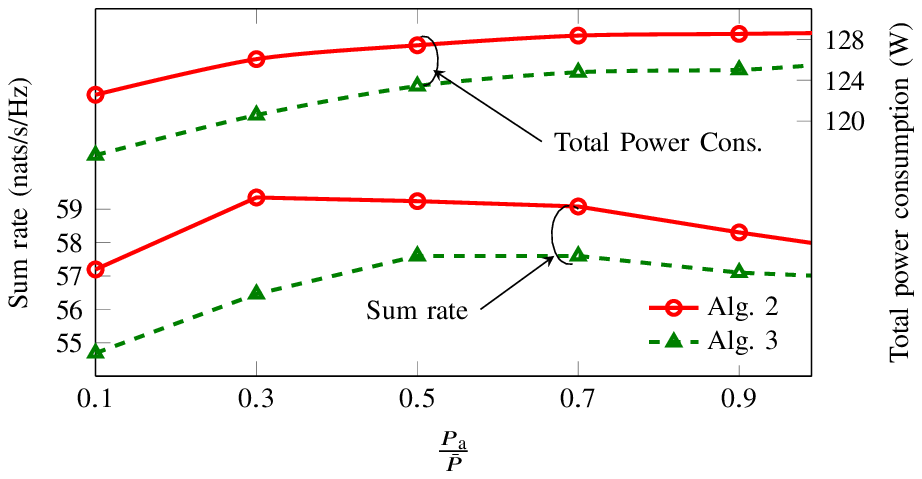}}
\par\end{centering}
\caption{Achieved performances versus the ratio $\frac{P_{\text{a}}}{\bar{P}}$
with $B=7$, $K=14$, $\bar{C}=40$ nats/s/Hz, $p_{\text{SP}}=10$
W/(Gnats/Hz) and $\bar{P}=30$ dBm.}
\label{Fig.9}
\end{figure}

In the final experiment, we fix $\bar{P}=30$ dBm and let $P_{\text{a}}$
vary to investigate the impact of the dynamics of PA's efficiency
on the EE performance. We recall that with some given $\epsilon_{\max}$
and input power, the PA's efficiency is a decreasing function with
respect to $P_{\text{a}}$ (see \eqref{eq:nonlinearpoweff}). Fig.
\ref{Fig.9a} plots the EE performances of Algorithms \ref{Alg.SCA}
and \algnew~versus $P_{\text{a}}$. The corresponding sum rate and
consumed power are shown in Fig. \ref{Fig.9b}. As can be seen from
Fig. \ref{Fig.9a}, when $P_{\text{a}}$ increases, the EE performance
first increases and then decreases. This observation can be explained
as follows. In the small regime of $P_{\text{a}}$, the transmit power
is small and an increase in the transmit power results in a significant
increase in the data rate, due to the logarithmic behavior of the
data rate w.r.t the transmit power. For this situation, the PA's efficiency
is still sufficiently high. Therefore, as $P_{\text{a}}$ increases,
the additional transmit power increases the sum rate more significantly
than the power consumption on PAs, and thus, it achieves better EE
(as can be seen from Fig. \ref{Fig.9b}). However, after a certain
value of $P_{\text{a}}$, the effective PA's efficiency becomes small
and its negative impact outweighs the benefit of increasing transmit
power. In this case, the reduced PA's efficiency due to the increase
of $P_{\text{a}}$ degrades EE performance.

\section{Conclusion}

This paper has studied the joint designs of beamforming, RRH-user
association and RRH selection in C-RANs to maximize the system EE
subject to per-RRH fronthaul capacity, transmit power budget and per-user
QoS. Specially, we have adopted relatively realistic power consumption
model compared to the previous works where the impacts of rate-dependent
signal processing power and the dynamics of PA's efficiency are considered.
To investigate the optimal performance of the formulated problem,
we have developed the new globally optimal method by customizing the
DBRB algorithm. We have also proposed novel modifications on the generic
framework of the DBRB method to improve the optimal algorithm's efficiency.
Towards practically appealing methods, we have proposed two suboptimal
approaches which can achieve very close to optimal performance with
much reduced complexity. Numerical evaluations have been provided
to demonstrate the effectiveness of the proposed schemes. Specifically,
the proposed modifications made on the DBRB framework remarkably reduce
the complexity of the globally optimal method. On the other hand,
the two proposed suboptimal approaches can achieve a near-optimal
solutions with a reasonable complexity and outperform the other known
methods. The impacts of the limited fronthaul capacity, rate-dependent
signal processing power and the dynamic of PA's efficiency on the
EE performance have also been demonstrated.

\section*{Appendix}

\subsection*{A. Proof of Lemma \ref{lem:1}}

First we show that \eqref{eq:achievable rate} is active, which is
proved by the contradiction. Let $(\eta^{\ast},{\bf w}^{\ast},{\bf r}^{\ast},{\bf t}^{\ast})$
be an optimal solution of \eqref{Prob:EE:epi} and suppose that \eqref{eq:achievable rate}
is not active at the optimum, i.e., $r_{k}^{\ast}<\log(1+\gamma_{k}({\bf w}^{\ast}))$
for some $k$. Then we can scale down the transmit power for user
$k$, i.e., $\|{\bf w}_{k}\|_{2}^{2}$, to achieve a new beamformer
$\|\hat{{\bf w}}_{k}\|_{2}^{2}$ such that $\|\hat{{\bf w}}_{k}\|_{2}^{2}=\tau\|{\bf w}_{k}\|_{2}^{2}<\|{\bf w}_{k}\|_{2}^{2}$
for $\tau\in(0,1)$ while keeping the others unchanged. By this way,
we can achieve $r_{k}^{\ast}<\log(1+\gamma_{k}(\hat{{\bf w}}))$ for
all $k$, since interference power at all users has reduced. However,
the new set of beamformers also generates a new power consumption
vector on PAs $\sum_{b\in{\cal B}}\hat{t}_{b}<\sum_{b\in{\cal B}}t_{b}^{\ast}$,
which immediately implies the increase of EE objective, i.e., $\eta>\eta^{\ast}$.
This contradicts to the fact that $(\eta^{\ast},{\bf w}^{\ast},{\bf r}^{\ast},{\bf t}^{\ast})$
is the optimal solution and thus completes the proof. Now for fixed
$({\bf x}^{\ast},{\bf s}^{\ast},{\bf r}^{\ast},{\bf t}^{\ast})$,
problem \eqref{Prob:Gen:Problem} reduces to a beamforming design
subject to the desired data rate ${\bf r}^{\ast}$ and the power constraint
$\sum_{i=1}^{I}||\tilde{\mathbf{w}}_{b,i}^{\ast}||_{2}=t_{b}^{\ast}$
such that at the output of \eqref{eq:find beam}, \eqref{eq:achievable rate}
must be binding.

\subsection*{B. Convergence Analysis of Algorithm 3}

We justify the convergence of Algorithm 3 by showing the following
facts: (i) when $\beta^{n}<\beta_{\max}$, the update of $\beta$
(see Step 5) tightens the approximations \eqref{eq:BS-user:epi:approx}
and \eqref{eq:BS:epi:approx} after every iteration; and (ii) let
$\bar{n}$ be the iteration such that $\beta^{\bar{n}-1}<\beta_{\max}$
and $\beta^{\bar{n}}=\beta_{\max}$, then we have the sequence $\{\eta^{n}\}_{n>\bar{n}}$
being non-decreasing, which is guaranteed to converge.

To prove (i), let us consider the non-smooth constraint \eqref{eq:BS-user:epi:approx},
i.e., $\mu_{b,k}\geq\bar{\varphi}_{\beta}(v_{b,k};v_{b,k}^{n})$ with
arbitrary $\beta$. Since it holds that $\bar{\varphi}_{\bar{\beta}}(.)\geq\bar{\varphi}_{\beta}(.)$
for any $\bar{\beta}\geq\beta$, we can replace $\bar{\varphi}_{\beta}(v_{b,k};v_{b,k}^{n})$
by $\bar{\varphi}_{\bar{\beta}}(v_{b,k};v_{b,k}^{n})$ in \eqref{eq:BS-user:epi:approx}
to obtain a tighter approximation, i.e., $\mu_{b,k}\geq\bar{\varphi}_{\bar{\beta}}(v_{b,k};v_{b,k}^{n})$.
Similarly, we can use the same argument for \eqref{eq:BS:epi:approx}.

Next, we prove (ii). We recall that the feasible set of \eqref{Prob:EE:sparsity:approx}
is bounded by power, fronthaul and users' QoS constraints. Thus it
is sufficient to prove that solution of \eqref{Prob:EE:sparsity:approx}
returned at iteration $n$ (i.e., $\tilde{\bm{\Omega}}^{n}$) is feasible
to the problem at iteration $n+1$ for $n>\bar{n}$, as such we yield
$\eta^{n+1}\geq\eta^{n}$ \cite{Giang:15:JCOML,Oskari:EE-optimalBeamDesign:14:JSP}.
To this end we note that $\tilde{\bm{\Omega}}^{n}$ satisfies (smooth)
constraints \eqref{eq:EE:epi:approx}\textendash \eqref{eq:BH:constraint:epi:approx}
at iteration $n+1$. This fact follows from the properties of convex
approximation which is also discussed in \cite[Properties (i) and (ii)]{MarksWright:78:AGenInnerApprox}.
On the other hand, for non-smooth constraint \eqref{eq:BS-user:epi:approx},
we have $\bar{\varphi}_{\beta_{\max}}(v_{b,k}^{n};v_{b,k}^{n})=\min\{1,\beta_{\max}v_{b,k}^{n}\}\leq\mu_{b,k}^{n}$.
This is because $(v_{b,k}^{n},\mu_{b,k}^{n})$ is the solution of
\eqref{Prob:EE:sparsity:approx} at iteration $n$. The result for
\eqref{eq:BS:epi:approx} can be obtained following the same manner.
At this point, we accomplish the argument (ii).

\bibliographystyle{IEEEtran}
\bibliography{IEEEabrv}

\begin{thebibliography}{10}
\providecommand{\url}[1]{#1}
\csname url@samestyle\endcsname
\providecommand{\newblock}{\relax}
\providecommand{\bibinfo}[2]{#2}
\providecommand{\BIBentrySTDinterwordspacing}{\spaceskip=0pt\relax}
\providecommand{\BIBentryALTinterwordstretchfactor}{4}
\providecommand{\BIBentryALTinterwordspacing}{\spaceskip=\fontdimen2\font plus
\BIBentryALTinterwordstretchfactor\fontdimen3\font minus
  \fontdimen4\font\relax}
\providecommand{\BIBforeignlanguage}[2]{{%
\expandafter\ifx\csname l@#1\endcsname\relax
\typeout{** WARNING: IEEEtran.bst: No hyphenation pattern has been}%
\typeout{** loaded for the language `#1'. Using the pattern for}%
\typeout{** the default language instead.}%
\else
\language=\csname l@#1\endcsname
\fi
#2}}
\providecommand{\BIBdecl}{\relax}
\BIBdecl

\bibitem{marsch2011coordinated}
P.~Marsch and G.~P. Fettweis, \emph{Coordinated Multi-Point in Mobile
  Communications: from theory to practice}.\hskip 1em plus 0.5em minus
  0.4em\relax Cambridge University Press, 2011.

\bibitem{3gppCoMP}
\BIBentryALTinterwordspacing
3GPP, ``Coordinated multi-point operation for {LTE} physical layer aspects
  {(Release 11)},'' 3rd Generation Partnership Project, TR 36.819. [Online].
  Available: \url{http://www.3gpp.org/technologies}
\BIBentrySTDinterwordspacing

\bibitem{mobile2011c}
C.~Mobile, ``{C-RAN}: the road towards green {RAN},'' \emph{White Paper},
  vol.~2, 2011.

\bibitem{Wu:2012:GreenWirelessComm}
J.~Wu, ``Green wireless communications: from concept to reality [industry
  perspectives],'' \emph{{IEEE} Wireless Commun.}, vol.~19, no.~4, pp. 4--5,
  Aug. 2012.

\bibitem{Checko:2014:CRANRevew}
A.~Checko, H.~L. Christiansen, Y.~Yan, L.~Scolari, G.~Kardaras, M.~S. Berger,
  and L.~Dittmann, ``Cloud ran for mobile networks -- {A} technology
  overview,'' \emph{{IEEE} Commun. Surveys Tuts.}, vol.~17, no.~1, pp.
  405--426, Firstquarter 2015.

\bibitem{Peng:2015:FHInsightChallenge}
M.~Peng, C.~Wang, V.~Lau, and H.~V. Poor, ``Fronthaul-constrained cloud radio
  access networks: insights and challenges,'' \emph{{IEEE} Wireless Commun.},
  vol.~22, no.~2, pp. 152--160, Apr. 2015.

\bibitem{Peng:ChanEst:Survey}
M.~Peng, Y.~Sun, X.~Li, Z.~Mao, and C.~Wang, ``Recent advances in cloud radio
  access networks: System architectures, key techniques, and open issues,''
  \emph{{IEEE} Commun. Surveys Tuts.}, vol.~18, no.~3, pp. 2282--2308,
  thirdquarter 2016.

\bibitem{FengJiang:13:aSurveyofEE}
D.~Feng, C.~Jiang, G.~Lim, J.~Cimini, L.~J., G.~Feng, and G.~Li, ``A survey of
  energy-efficient wireless communication,'' \emph{{IEEE} Commun. Surveys
  Tuts.}, vol.~15, no.~1, pp. 167--178, Feb. 2013.

\bibitem{zappone2015energy}
A.~Zappone and E.~Jorswieck, ``Energy efficiency in wireless networks via
  fractional programming theory,'' \emph{Foundations and Trends in
  Communications and Information Theory}, vol.~11, no. 3-4, pp. 185--396, 2015.

\bibitem{Dan:2013:SP:Fullduplex}
D.~Nguyen, L.-N. Tran, P.~Pirinen, and M.~Latva-aho, ``Precoding for full
  duplex multiuser {MIMO} systems: {S}pectral and energy efficiency
  maximization,'' \emph{{IEEE} Trans. Signal Process.}, vol.~61, no.~16, pp.
  4038--3050, Aug. 2013.

\bibitem{DerrickKwanNg:2012:JWCOM:EE_OFDM}
D.~W.~K. Ng, E.~S. Lo, and R.~Schober, ``Energy-efficient resource allocation
  in multi-cell {OFDMA} systems with limited backhaul capacity,'' \emph{{IEEE}
  Trans. Wireless Commun.}, vol.~11, no.~10, pp. 3618--3631, Oct. 2012.

\bibitem{HHJY:13:JCOM}
S.~He, Y.~Huang, S.~Jin, and L.~Yang, ``Coordinated beamforming for energy
  efficient transmission in multicell multiuser systems,'' \emph{{IEEE} Trans.
  Commun.}, vol.~61, no.~12, pp. 4961--4971, Dec. 2013.

\bibitem{Giang:15:JCOML}
K.-G. Nguyen, L.-N. Tran, O.~Tervo, Q.-D. Vu, and M.~Juntti, ``Achieving energy
  efficiency fairness in multicell multiuser {MISO} downlink,'' \emph{{IEEE}
  Commun. Lett.}, vol.~19, no.~8, pp. 1426 -- 1429, Aug. 2015.

\bibitem{Oskari:EE-optimalBeamDesign:14:JSP}
O.~Tervo, L.-N. Tran, and M.~Juntti, ``Optimal energy-efficient transmit
  beamforming for multi-user {MISO} downlink,'' \emph{{IEEE} Trans. Signal
  Process.}, vol.~63, no.~20, pp. 5574 -- 5588, Oct. 2015.

\bibitem{zappone:2017:GlobalEE}
A.~Zappone, E.~Bj\"{o}rnson, L.~Sanguinetti, and E.~Jorswieck, ``Globally
  optimal energy-efficient power control and receiver design in wireless
  networks,'' \emph{{IEEE} Trans. Signal Process.}, vol.~65, no.~11, pp.
  2844--2859, Jun. 2017.

\bibitem{HHJYL:13:JCOML}
S.~He, Y.~Huang, S.~Jin, F.~Yu, and L.~Yang, ``Max-min energy efficient
  beamforming for multicell multiuser joint transmission systems,''
  \emph{{IEEE} Commun. Lett.}, vol.~17, no.~10, pp. 1956--1959, Oct. 2013.

\bibitem{Isheden:2010:globecom}
C.~Isheden and G.~P. Fettweis, ``Energy-efficient multi-carrier link adaptation
  with sum rate-dependent circuit power,'' in \emph{2010 IEEE GLOBECOM 2010},
  Dec. 2010, pp. 1--6.

\bibitem{Xiong:2012:EEOFDMA}
C.~Xiong, G.~Y. Li, S.~Zhang, Y.~Chen, and S.~Xu, ``Energy-efficient resource
  allocation in {OFDMA} networks,'' \emph{{IEEE} Trans. Wireless Commun.},
  vol.~60, no.~12, pp. 3767--3778, Dec. 2012.

\bibitem{Wang:2013:EERate-dependentPower}
T.~Wang and L.~Vandendorpe, ``On the optimum energy efficiency for flat-fading
  channels with rate-dependent circuit power,'' \emph{{IEEE} Trans. Commun.},
  vol.~61, no.~12, pp. 4910--4921, Dec. 2013.

\bibitem{Tran:2017:VirtualizedBBU}
T.~X. Tran, A.~Younis, and D.~Pompili, ``Understanding the computational
  requirements of virtualized baseband units using a programmable cloud radio
  access network testbed,'' in \emph{2017 IEEE International Conference on
  Autonomic Computing (ICAC)}, Columbus, Ohio,USA, July 2017, pp. 221--226.

\bibitem{Hajisami:2017:ElasticNet}
A.~Hajisami, T.~X. Tran, and D.~Pompili, ``Elastic-net: Boosting energy
  efficiency and resource utilization in {5G C-RANs},'' in \emph{2017 IEEE 14th
  International Conference on Mobile Ad Hoc and Sensor Systems (MASS)},
  Orlando, FL, USA, Oct 2017, pp. 466--470.

\bibitem{Mikami2007}
S.~Mikami, T.~Takeuchi, H.~Kawaguchi, C.~Ohta, and M.~Yoshimoto, ``An
  efficiency degradation model of power amplifier and the impact against
  transmission power control for wireless sensor networks,'' in \emph{2007 IEEE
  Radio and Wireless Symposium,}, 2007, pp. 447--450.

\bibitem{Bjoernemo2009}
E.~Bj{\"o}rnemo, ``Energy constrained wireless sensor networks: Communication
  principles and sensing aspects,'' Ph.D. dissertation, Institutionen f{\"o}r
  teknikvetenskaper, 2009.

\bibitem{AmplifierMIMO-Persson}
D.~Persson, T.~Eriksson, and E.~G. Larsson, ``Amplifier-aware multiple-input
  multiple-output power allocation,'' \emph{{IEEE} Commun. Lett.}, vol.~17,
  no.~6, pp. 1112--1115, Jun. 2013.

\bibitem{BinBinDai:15:Access:BeamformingBH}
B.~Dai and W.~Yu, ``Sparse beamforming and user-centric clustering for downlink
  cloud radio access network,'' \emph{IEEE Access}, vol.~2, no.~6, pp.
  1326--1339, Oct. 2014.

\bibitem{Dai2016:EECRAN}
------, ``Energy efficiency of downlink transmission strategies for cloud radio
  access networks,'' \emph{{IEEE} J. Sel. Areas Commun.}, vol.~34, no.~4, pp.
  1037--1050, April 2016.

\bibitem{Tran:2017:DynamicCRAN}
T.~X. Tran and D.~Pompili, ``Dynamic radio cooperation for user-centric
  {cloud-RAN} with computing resource sharing,'' \emph{{IEEE} Trans. Wireless
  Commun.}, vol.~16, no.~4, pp. 2379--2393, Apr. 2017.

\bibitem{FuxinKLau:2014:SP:BackhaulLimitedSDPRelax}
F.~Zhuang and V.~K.~N. Lau, ``Backhaul limited asymmetric cooperation for
  {MIMO} cellular networks via semidefinite relaxation,'' \emph{{IEEE} Trans.
  Signal Process.}, vol.~62, no.~3, pp. 684--693, Feb. 2014.

\bibitem{GiangICASSP2017}
K.~G. Nguyen, Q.~D. Vu, M.~Juntti, and L.~N. Tran, ``Globally optimal
  beamforming design for downlink {CoMP} transmission with limited backhaul
  capacity,'' in \emph{2017 IEEE International Conference on Acoustics, Speech
  and Signal Processing (ICASSP)}, March 2017.

\bibitem{HAJISAMIDJP}
\BIBentryALTinterwordspacing
A.~Hajisami and D.~Pompili, ``Dynamic joint processing: Achieving high spectral
  efficiency in uplink {5G} cellular networks,'' \emph{Computer Networks}, vol.
  126, pp. 44 -- 56, 2017. [Online]. Available:
  \url{http://www.sciencedirect.com/science/article/pii/S1389128617302700}
\BIBentrySTDinterwordspacing

\bibitem{HAJISAMICFFR}
------, ``Joint virtual edge-clustering and spectrum allocation scheme for
  uplink interference mitigation in {C-RAN},'' \emph{Ad Hoc Networks}, vol.~72,
  pp. 91 -- 104, 2018.

\bibitem{GroupGreenBeam-YShi}
Y.~Shi, J.~Zhang, and K.~B. Letaief, ``Group sparse beamforming for green
  {cloud-RAN},'' \emph{{IEEE} Trans. Wireless Commun.}, vol.~13, no.~5, pp.
  2809--2823, May 2014.

\bibitem{tuy2006discrete}
H.~Tuy, M.~Minoux, and N.~Hoai-Phuong, ``Discrete monotonic optimization with
  application to a discrete location problem,'' \emph{SIAM Journal on
  Optimization}, vol.~17, no.~1, pp. 78--97, 2006.

\bibitem{le2015dc}
H.~A. Le~Thi, T.~P. Dinh, H.~M. Le, and X.~T. Vo, ``{DC} approximation
  approaches for sparse optimization,'' \emph{European Journal of Operational
  Research}, vol. 244, no.~1, pp. 26--46, 2015.

\bibitem{HowmuchEnergy-Auer}
G.~Auer, V.~Giannini, C.~Desset, I.~Godor, P.~Skillermark, M.~Olsson, M.~A.
  Imran, D.~Sabella, M.~J. Gonzalez, O.~Blume, and A.~Fehske, ``How much energy
  is needed to run a wireless network?'' \emph{{IEEE} Wireless Commun.},
  vol.~18, no.~5, pp. 40--49, Oct. 2011.

\bibitem{Dhaini:2014:EE-TDMA}
A.~R. Dhaini, P.~H. Ho, G.~Shen, and B.~Shihada, ``Energy efficiency in
  {TDMA-Based} next-generation passive optical access networks,''
  \emph{{IEEE/ACM} Trans. Netw.}, vol.~22, no.~3, pp. 850--863, Jun. 2014.

\bibitem{tuy2005monotonic}
H.~Tuy, F.~Al-Khayyal, and P.~T. Thach, ``Monotonic optimization: Branch and
  cut methods,'' in \emph{Essays and Surveys in Global Optimization}.\hskip 1em
  plus 0.5em minus 0.4em\relax Springer, 2005, pp. 39--78.

\bibitem{Bjornson2012RobustMonotonicMISO}
E.~Bj\"{o}rnson, G.~Zheng, M.~Bengtsson, and B.~Ottersten, ``Robust monotonic
  optimization framework for multicell {MISO} systems,'' \emph{{IEEE} Trans.
  Signal Process.}, vol.~60, no.~5, pp. 2508--2523, May 2012.

\bibitem{Jorswieck2010MonotonicMISO}
E.~A. Jorswieck and E.~G. Larsson, ``Monotonic optimization framework for the
  two-user {MISO} interference channel,'' \emph{{IEEE} Trans. Commun.},
  vol.~58, no.~7, pp. 2159--2168, Jul. 2010.

\bibitem{PhuongTSP2017}
P.~Luong, F.~Gagnon, C.~Despins, and L.~N. Tran, ``Optimal joint remote radio
  head selection and beamforming design for limited fronthaul {C-RAN},''
  \emph{{IEEE} Trans. Signal Process.}, vol.~65, no.~21, pp. 5605--5620, Nov.
  2017.

\bibitem{dattorro2010convex}
J.~Dattorro, \emph{Convex optimization \& Euclidean distance geometry}.\hskip
  1em plus 0.5em minus 0.4em\relax Lulu. com, 2010.

\bibitem{MarksWright:78:AGenInnerApprox}
B.~R. Marks and G.~P. Wright, ``A general inner approximation algorithm for
  nonconvex mathematical programs,'' \emph{Operations Research}, vol.~26,
  no.~4, pp. 681--683, Jul.-Aug. 1978.

\bibitem{le2014dc}
H.~A. Le~Thi, T.~P. Dinh \emph{et~al.}, ``{DC} programming and {DCA} for
  general {DC} programs,'' in \emph{Advanced Computational Methods for
  Knowledge Engineering}.\hskip 1em plus 0.5em minus 0.4em\relax Springer,
  2014, pp. 15--35.

\bibitem{le2012exact}
H.~A. Le~Thi, T.~Pham~Dinh, and H.~V. Ngai, ``Exact penalty and error bounds in
  {DC} programming,'' \emph{Journal of Global Optimization}, vol.~52, no.~3,
  pp. 509--535, 2012.

\bibitem{zhang2010analysis}
T.~Zhang, ``Analysis of multi-stage convex relaxation for sparse
  regularization,'' \emph{Journal of Machine Learning Research}, vol.~11, no.
  Mar, pp. 1081--1107, 2010.

\bibitem{BenNemi:book:LectModConv}
A.~Ben-Tal and A.~Nemirovski, \emph{Lectures on modern convex
  optimization}.\hskip 1em plus 0.5em minus 0.4em\relax Philadelphia: {MPS-SIAM
  Series on Optimization, SIAM}, 2001.

\bibitem{Mosek}
{I. MOSEK ApS, 2014}, {[Online]. Available: www.mosek.com}.

\end{thebibliography}

\end{document}